\pdfoutput=1
\documentclass[fleqn,usenatbib]{mnras}

\usepackage{newtxtext,newtxmath}

\usepackage[T1]{fontenc}

\DeclareRobustCommand{\VAN}[3]{#2}
\let\VANthebibliography\thebibliography
\def\thebibliography{\DeclareRobustCommand{\VAN}[3]{##3}\VANthebibliography}

\usepackage{graphicx}	
\usepackage{amsmath}	

\title[Jet in NGC~2617]{A compact core-jet structure in the changing-look Seyfert NGC~2617}

\author[J. Yang et al.]{Jun Yang,$^{1,2}$\thanks{E-mail: jun.yang@chalmers.se}, 
Zsolt Paragi,$^{2}$
Robert J. Beswick$^{3}$,
Wen Chen,$^{4}$ 
Ilse M. van Bemmel,$^{2}$
Qingwen Wu,$^{5}$
\and
Tao An,$^{6}$
Xiaocong Wu,$^{6}$
Lulu Fan,$^{7,8,9}$
J.~B.~R. Oonk,$^{10,11,12}$ 
Xiang Liu$^{13}$
and
Weihua Wang$^{14}$
\\
$^{1}$Department of Space, Earth and Environment, Chalmers University of Technology, Onsala Space Observatory, SE-439 92 Onsala, Sweden \\
$^{2}$Joint Institute for VLBI ERIC (JIVE), Postbus 2, NL-7990 AA Dwingeloo, the Netherlands \\
$^{3}$Jodrell Bank Centre for Astrophysics, School of Physics and Astronomy, The
University of Manchester, Manchester M13 9PL, UK \\
$^{4}$Yunnan Observatories, Chinese Academy of Sciences, 396 Yangfangwang, Guandu District, Kunming, 650216, P. R. China \\
$^{5}$Department of Astronomy, School of Physics, Huazhong University of Science and Technology, Wuhan 430074, China \\
$^{6}$Shanghai Astronomical Observatory, Key Laboratory of Radio Astronomy, Chinese Academy of Sciences, 200030 Shanghai, China \\
$^{7}$CAS Key Laboratory for Research in Galaxies and Cosmology, Department of Astronomy, University of Science and Technology of China, Hefei 230026, China \\
$^{8}$School of Astronomy and Space Sciences, University of Science and Technology of China, Hefei 230026, China \\
$^{9}$Shandong Provincial Key Lab of Optical Astronomy and Solar-Terrestrial Environment, Institute of Space Science, Shandong University, Weihai 264209, China \\
$^{10}$SURFsara, PO Box 94613, NL-1090 GP Amsterdam, the Netherlands \\
$^{11}$Leiden Observatory, Leiden University, PO Box 9513, NL-2300 RA Leiden, the Netherlands \\
$^{12}$Netherlands Institute for Radio Astronomy (ASTRON), NL-7991 PD Dwingeloo, the Netherlands \\
$^{13}$Xinjiang Astronomical Observatory, Key Laboratory of Radio Astronomy, Chinese Academy of Sciences,150 Science 1-Street, 830011 Urumqi, China \\
$^{14}$School of Computer Information Engineering, Changzhou Institute of Technology, Changzhou 213032, China \\
}

\date{Accepted XXX. Received YYY; in original form ZZZ}

\pubyear{2021}

\begin{document}
\label{firstpage}
\pagerange{\pageref{firstpage}--\pageref{lastpage}}
\maketitle
\begin{abstract}
The nearby face-on spiral galaxy NGC\,2617 underwent an unambiguous `inside-out' multi-wavelength outburst in Spring 2013, and a dramatic Seyfert type change probably between 2010 and 2012, with the emergence of broad optical emission lines. To search for the jet activity associated with this variable accretion activity, we carried out multi-resolution and multi-wavelength radio observations. Using the very long baseline interferometric (VLBI) observations with the European VLBI Network (EVN) at 1.7 and 5.0~GHz, we find that NGC~2617 shows a partially synchrotron self-absorbed compact radio core with a significant core shift, and an optically thin steep-spectrum jet extending towards the north up to about two parsecs in projection. We also observed NGC~2617 with the electronic Multi-Element Remotely Linked Interferometer Network (\textit{e}-MERLIN) at 1.5 and 5.5~GHz, and revisited the archival data of the Very Large Array (VLA) and the Very Long Baseline Array (VLBA). The radio core had a stable flux density of $\sim$1.4~mJy at 5.0~GHz between 2013 June and 2014 January, in agreement with the expectation of a supermassive black hole in the low accretion rate state. The northern jet component is unlikely to be associated with the `inside-out' outburst of 2013. Moreover, we report that most optically selected changing-look AGN at $z<0.83$ are sub-mJy radio sources in the existing VLA surveys at 1.4~GHz, and it is unlikely that they are more active than normal AGN at radio frequencies. 
\end{abstract}

\begin{keywords}
galaxies: active -- galaxies: individual: NGC~2617 -- galaxies: jets -- galaxies: Seyfert -- radio continuum: galaxies
\end{keywords}



\section{Introduction}
\label{sec:intro}
The broad-line and continuum components of active galactic nuclei (AGN) may vary moderately \citep[e.g. NGC~4151,][]{Penston1984}. In the extreme cases, broad Balmer emission lines could vanish completely from type 1 galaxies or appear from type 2 galaxies.  Recently, well-observed and extreme cases include the Seyfert galaxies NGC~2617 \citep{Shappee2014, Oknyansky2017}, Mrk~590 \citep{Denney2014, Koay2016VLBA, Yang2021}, Mrk~1018 \citep{McElroy2016, Noda2018}, X-ray AGN 1ES~1927$+$654 \citep[e.g.][]{Gabanyi2014, Trakhtenbrot2019, Ricci2020} and the quasar SDSS~J015957.64$+$003310.5 \citep{LaMassa2015}. Moreover, the number of changing-look AGN has recently been significantly increased by some dedicated search campaigns \citep[e.g. ][]{Runco2016, Yang2018, Frederick2019, MacLeod2019, Guo2020}.

These dramatic type changes are generally interpreted as a consequence of variable accretion activity \citep[e.g.][]{Peterson1986, LaMassa2015, Noda2018} rather than variable obscuring material along the line of sight. On the one hand, the scenario of the dust obscuration fails to naturally explain large variations of the mid-infrared luminosity \citep[10 changing-look AGN,][]{Sheng2017} and low linear polarisation degrees in low ultra-violet and optical blue bands \citep[13 changing-look quasars,][]{Hutsemkers2019}. On the other hand, a high increase in accretion rate can power stronger broad lines \citep[e.g.][]{Peterson1986, Elitzur2014}. There is also a positive correlation ($R_{\rm BLR} \propto L^\rho$, $\rho$ in the range of 0.56--0.70) between the characteristic broad-line region size $R_{\rm BLR}$ and the Balmer emission line, X-ray, ultraviolet (UV), and optical continuum luminosity $L$ \citep{Kaspi2005}. If extreme changing-look phenomena represent some short, intensive, and probably episodic accretion activity of supermassive black holes (SMBHs) in the very inner accretion disks, they would be highly interesting objects for radio observations to search for short-lived jets on parsec scales \citep{Woowska2017, Yang2021} and to probe complex accretion-ejection activity \citep[e.g.][]{Marscher2002, Fender2009, Yuan2014, Blandford2019}.

The nearby face-on galaxy NGC~2617 at $z = 0.0142$ \citep{Paturel2003} was first identified by \citet{Moran1996} as a Seyfert~1.8 spiral galaxy because of very weak broad lines. \citet{Shappee2014} reported that its central AGN had a strong multi-band AGN outburst between 2013 April 24 and June 20 from X-ray to near-infrared (NIR) wavelengths with increasing time lags up to $\sim$8~days, which can be naturally explained by the reprocessing of the inner X-ray emission. It showed a dramatic Seyfert type change from 1.8 to 1, with the appearance of broad optical emission lines. \citet{Oknyansky2017} proposed that the Seyfert type change probably occurred between 2010 October and 2012 February. 

NGC~2617 also has a radio counterpart first detected in the NRAO (National Radio Astronomical Observatory) VLA Sky Survey \citep[NVSS,][]{Condon1998NVSS}. Its host galaxy shows significant radio emission of HI line in the HI Parkes All Sky Survey (HIPASS) catalogue \citep{Doyle2005}. To search for short-lived, parsec-scale jets associated with the outburst and type changes, we initiated VLBI imaging observations of NGC~2617. The pilot high-resolution VLBI observations found a compact structure \citep{Yang2013} and a flat spectrum between 1.7 and 5.0~GHz \citep{Jencson2013}. The data of these pilot observations are also re-visited by us in the paper. The updated results are reported in Table~\ref{tab:flux}. Moreover, we carried out follow-up dual-frequency EVN and \textit{e}-MERLIN observations. 

\begin{table*}
\caption{List of the radio flux densities of NGC~2617. Columns give (1) observing band, (2) date, (3) integrated flux density and total error including the systematic error (five per cent for the VLA and \textit{e}-MERLIN observations, ten per cent for the VLBI observations), (4) map peak brightness and statistical off-source root mean square, (5) observing frequency (6) major and minor axes, position angle of synthesised beam, (7) used array (and configuration), (8) project name and (9) reference.  }
\label{tab:flux}
\begin{tabular}{ccccccccc} 
\hline
Band 
  & Date         & $S_{\rm int}$ & $S_{\rm pk}$ & $\nu_{\rm obs}$ 
                                                         & Synthesised Beam                       &  Array &  Project & Reference  \\
  &              & (mJy) & (mJy\,beam$^{-1}$)    & (GHz) & (Major, Minor, PA)                     &        &          & \\
\hline
L & 1993 Apr 26  & $19.8\pm1.0$  & $13.6\pm0.1$  & 1.40  & $5.2\times4.5$ arcsec$^2$, $+12\fdg1$  &  VLA:B & AT0149B  &  This paper \\ 
L & 1993 Jul 06  & $24.5\pm1.2$  & $11.2\pm0.1$  & 1.43  & $29.4\times16.1$ arcsec$^2$, $+34\fdg1$&  VLA:D & AT0149D  &  This paper \\ 
L & 1993 Nov 15  & $28.1\pm1.6$  & $21.2\pm0.5$  & 1.40  & $45.0\times45.0$ arcsec$^2$            &  VLA:D & NVSS     & \citet{Condon1998NVSS} \\ 
L & 1995 Jul 10  & $4.8\pm0.3$   & $3.2\pm0.3$   & 1.43  & $2.0\times1.4$ arcsec$^2$, $-18\fdg7$  &  VLA:A & AM0492A  & This paper \\
\hline
L & 2013 Jun 07  & $1.2\pm0.3$   & $0.82\pm0.05$ & 1.67  &   $24.7\times3.7$ mas$^{2}$, $+74\fdg5$ &  EVN  & RY005   & This paper \\
L & 2013 Jun 29  & $1.3\pm0.2$   & $1.25\pm0.10$ & 1.64  &  $11.9\times5.0$ mas$^2$, $+4\fdg4$    & VLBA  & SS004   & This paper \\
L & 2013 Sep 18  & $1.4\pm0.1$   & $0.85\pm0.03$ & 1.66  &  $24.0\times3.2$ mas$^{2}$, $+77\fdg0$ & EVN    & EY021A  & This paper \\
L & 2014 Jan 25  & $2.0\pm0.2$   & $2.15\pm0.18$ & 1.53  &   $623.0\times109.0$~mas$^{2}$, $+19\fdg$9 & $e$-MERLIN   & CY1026   & This paper \\
\hline
C & 2013 Jun 29  & $1.4\pm0.1$   & $1.28\pm0.07$ & 4.99  &  $3.8\times1.6$ mas$^2$, $+2\fdg2$     & VLBA   & SS004   & This paper \\
C & 2013 Oct 09  & $1.2\pm0.1$   & $1.10\pm0.02$ & 4.99  &   $6.3\times2.0$~mas$^{2}$, $+76\fdg$9 & EVN    & EY021B   & This paper \\
C & 2014 Jan 11  & $1.5\pm0.2$   & $1.51\pm0.11$ & 5.50  &   $104.0\times71.8$~mas$^{2}$, $+29\fdg$2 & $e$-MERLIN   & CY1026   & This paper \\
\hline
X & 1995 Jul 10  & $\leq0.3$ (3$\sigma$) & $\leq0.3 (3\sigma)$    
                                            & 8.44  & $0.33\times0.23$ arcsec$^2$, $-22\fdg3$ & VLA:A & AM0492A  & This paper \\
\hline
\end{tabular}
\end{table*}

The paper is organised in the following sequence. We introduce our radio observations and data reduction in Section~\ref{sec:obs}. We present our imaging results of NGC~2617 in Section~\ref{sec:results}. In Section~\ref{sec:discussion}, we interpret its parsec-scale radio morphology, investigate possible jet activity associated with the Seyfert type change and the outburst of 2013, and discuss short-lived radio sources and the population of optically selected changing-look AGN in general. Throughout the paper, a standard $\Lambda$CDM cosmological model with $H_{\rm 0}$~=~71~km\,s$^{-1}$\,Mpc$^{-1}$, $\Omega_{\rm m}$~=~0.27, $\Omega_{\Lambda}$~=~0.73 is adopted. The VLBI images of NGC~2617 have a scale of 0.31~pc\,mas$^{-1}$. The error propagation is done via the \textsc{python} package \textsc{uncertainty}\footnote{\url{https://pythonhosted.org/uncertainties/}}.

\section{Observations and data reduction}
\label{sec:obs}

\begin{table*}
\caption{Configurations of the EVN and \textit{e}-MERLIN observations of NGC~2617. The two-letter code for each station is explained in Section~\ref{sec:obs}.  }
\label{tab:exp}
\begin{tabular}{lcccllcc}
\hline
Project  & Freq. 
                 & Bandwidth 
                         & Starting time       & Duration &  Participating Stations                      & Phase-Referencing   \\    
 Code    & (GHz) & (MHz) & (UT)                &  (hr)    &                                              &  Quality            \\       
\hline     
 RY005   & 1.66  & 128   & 2013 Jun 07, 11h    &  7       & \texttt{EF, WB, JB1, HH, ON, NT, MC, TR}     & Poor                \\ 
EY021A   & 4.99  & 128   & 2013 Sep 18, 05h    &  7       & \texttt{EF, WB, JB2, HH, ON, NT, MC, TR, YS} & Successful          \\    
EY021B   & 1.66  & 128   & 2013 Oct 09, 03h    &  8       & \texttt{EF, WB, JB2, HH, ON, NT, MC}         & Successful          \\   
\hline
CY1026   & 5.50  & 512   & 2014 Jan 11, 21h    &  18      & \texttt{JB2, KN, PI, DA, CM, DE}             & Successful          \\
CY1026   & 1.53  & 512   & 2014 Jan 24, 20h    &  9       & \texttt{JB2, KN, PI, DA, CM}                 & Successful          \\
\hline
\end{tabular}
\end{table*}

\subsection{EVN observations at 1.7 and 5.0~GHz}
We observed NGC~2617 with the EVN at 1.7 and 5.0~GHz in 2013. The basic experiment parameters are listed in Table~\ref{tab:exp}.  The participating EVN stations were Effelsberg (\texttt{EF}), phased-up array of the Westerbork Synthesis Radio Telescope (\texttt{WB}),  Jodrell Bank Lovell (\texttt{JB1}) and MK II (\texttt{JB2}), Hartebeesthoek (\texttt{HH}), Onsala (\texttt{ON}), Noto (\texttt{NT}),  Medicina (\texttt{MC}), Toru\'n (\texttt{TR}) and Yebes (\texttt{YS}). The experiments used the available maximum data rate 1024~Mbps (16~MHz filters, 2~bit quantisation, 16~sub-bands in dual polarisation). All the three experiments were carried out in the $e$-VLBI mode \citep{Szomoru2008}. The data were streamed to JIVE (Joint Institute for VLBI ERIC) via broad-band fibre connections and then correlated in the real-time mode by the EVN software correlator \citep[\textsc{SFXC},][]{Keimpema2015} using the general correlation parameters (1 or 2 s integration time, 32 or 64 points per subband) for continuum experiments. 

We used J0834$-$0417 as the phase-referencing calibrator during the observations of NGC~2617. The calibrator has an angular separation of 16~arcmin to our target NGC~2617. The correlation position for the calibrator is RA~=~08$^{\rm h}$34$^{\rm m}$53$\fs$47,  Dec~=~$-$04$\degr$17$\arcmin$11$\farcs$4. In the phase-referencing astrometry, we took the more accurate position RA~=~~08$^{\rm h}$34$^{\rm m}$53$\fs$475012, DEC~=~$-$04$\degr$17$\arcmin$11$\farcs$36845 ($\sigma_{\rm ra} = 0.5$ mas, $\sigma_{\rm dec} = 1.2$ mas) provided by the radio fundamental catalogue (RFC) 2020B \citep{Petrov2021}. The cycle time is $\sim$4 minutes ($\sim$40~s for J0834$-$0417, $\sim$150~s for NGC~2617, $\sim$40~s for two gaps). We also inserted several short scans of bright calibrators B0845$-$051 and 4C~39.25.

The correlation output data were calibrated with the NRAO software package Astronomical Image Processing System \citep[{\sc aips},][]{Greisen2003}. When the visibility data were loaded into \textsc{aips} disks,  we excluded one quarter of side channels because of their very low correlation amplitude. In view of the data flagging, we re-normalised the auto-correlation and cross-correlation amplitudes of the data with the \textsc{aips} task \texttt{ACCOR}. Before a-priori amplitude calibration via antenna gain information, the antenna system temperature data were properly smoothed to minimise their noise. When antenna system temperatures and gain curves were not available, we used nominal system equivalent flux densities and a flat gain curve to do the amplitude calibration. The task \texttt{TECOR} was used to remove ionospheric dispersive delays calculated according to maps of total electron content provided by Global Positioning System (GPS) satellite observations. The phase errors associated with the antenna parallactic angle variations were removed. We corrected the instrumental phases and delays across the subbands via running fringe-fitting with a short scan of the calibrator data. After the precise phase alignment, all the subband data were combined to run fringe fitting with a solution interval of about one minute. The sensitive station (\texttt{EF} or \texttt{Wb}) was used as the reference station.  The solutions were also applied to all the related sources via a two-point linear interpolation. The bandpass calibration was performed. To run these related \textsc{aips} tasks in a script, we used the \textsc{ParselTongue} interface \citep{Kettenis2006}. 

\begin{figure}
\includegraphics[width=\columnwidth]{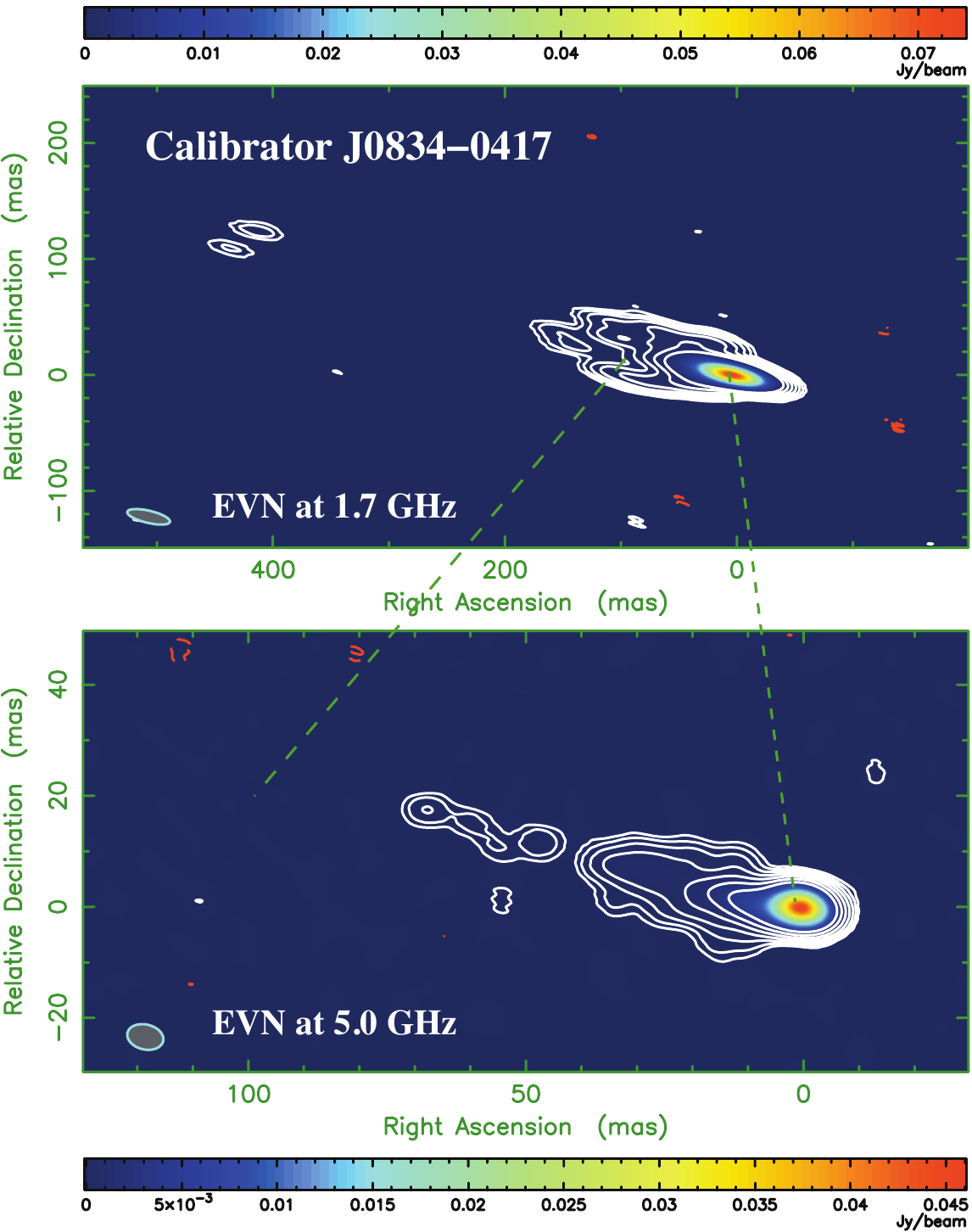}  \\
\caption{
The total intensity images of the calibrator J0834$-$0417. The contours start from 3$\sigma$ and increase by factors of $-$1, 1, 2, ..., 64. \textbf{Top}: The EVN image at 1.7~GHz. The restoring beam has a full width at half maximum (FWHM) of 37.8~$\times$~10.5~mas$^{2}$ at a position angle (PA) of $+76\fdg8$. The first contour is 0.027~mJy\,beam$^{-1}$ and the peak brightness is 74.1~mJy\,beam$^{-1}$. \textbf{Bottom}: The EVN image at 5.0~GHz. The beam FWHM is 6.7~$\times$~1.0~mas$^{2}$ at 79$\fdg$0. The first contour is 0.056~mJy\,beam$^{-1}$ (3$\sigma$) and the peak brightness is 46.1~mJy\,beam$^{-1}$.}
\label{fig:cal}
\end{figure}

The deconvolution was performed in \textsc{difmap} \citep{Shepherd1994}. The calibrator imaging procedure was performed through several iterations of model fitting with some delta functions, i.e., point source models, and the self-calibration in \textsc{difmap} \citep{Shepherd1994}. This special method would allow us to obtain the best possible dynamic range image \citep[e.g.][]{Yang2020imbh} in case of a poor ($u$, $v$) coverage and a complex VLBI network \citep[e.g.][]{Yang2021pds}. The calibrator J0834$-$0417 had total flux densities $\sim$96~mJy at 1.7~GHz, $\sim$66~mJy at 5.0~GHz, $\sim$96~mJy at 1.7~GHz over the three epochs. Fig.~\ref{fig:cal} shows the imaging results of the phase-referencing calibrator J0834$-$0417 in the last two experiments. The calibrator shows a one-sided core-jet structure at both 1.7 and 5.0~GHz. Although the ($u$, $v$) coverage is poor on the long baselines, the deconvolution method allows us to achieve a reasonable dynamic range. With the input images, we re-ran the fringe-fitting and the amplitude and phase self-calibration in \textsc{aips}. All these solutions were also transferred to the target data by the linear interpolation. 

To avoid bandwidth smearing effect, we used the \textsc{aips} tasks \texttt{MORIF} and \texttt{SPLIT} to rearrange the data structure of the target source NGC~2617. The output data had a relatively high frequency resolution of 4~MHz per intermediate frequency subband.  We imaged NGC~2617 with the \texttt{clean} algorithm in \textsc{difmap}. The data on the sensitive short baseline \texttt{EF--WB} were excluded to remove some stripes in the dirty map. Because of the strong side lobes resulting from the poor ($u$, $v$) coverage, the \texttt{clean} windows were carefully added and enlarged to avoid cleaning near or at the side lobes. Phase self-calibration was not used during the imaging process for NGC~2617. 

Our target source NGC~2617 was detected in all the three epochs. In the first epoch, the image had a significantly high image noise level, because the most sensitive station \texttt{EF} was not available for the first five hours. In addition, there were significant phase fluctuations on short timescales $\sim$5~min, most likely because of the unstable ionosphere resulting from unexpected intense solar activity during the daytime. This problem caused a poor deconvolution and a high noise level near the target source. Thus, we failed to detect any secondary component (e.g. J in Fig.~\ref{fig:ngc2617}) near the peak feature on 2013 June 7.   

\subsection{Broad-band \textit{e}-MERLIN observations}
We also observed NGC~2617 with the \textit{e}-MERLIN array at L and C bands. The C-band observations were done with the standard frequency setup (4 subbands, 128~MHz per subband, dual polarisation, 8-bit quantisation). The L band observations were performed with the default frequency setup (8 subbands, 64~MHz per subband, dual polarisation, 2-bit quantisation). The participating stations were \texttt{JB2}, Knocking (\texttt{KN}), Pickmere (\texttt{PI}), Darnhall (\texttt{DA}), Cambridge (\texttt{CM}) and Defford (\texttt{DE}).  During the L and C-band experiments, the flux calibrators were 3C~286 and OQ~208, the bandpass calibrator was 0319$+$415, and the phase-referencing calibrator was J0834$-$0417. 

The data were also calibrated in \textsc{aips}. First, we reviewed the data quality and flagged out very noisy data. Second, we did the phase and bandpass calibrations in the same way as the above EVN data reduction. Third, we set the flux density of the primary flux calibrator 3C~286, ran amplitude and phase self-calibration on 3C~286, using its standard VLA image as the input source model, and treat the other calibrators as point-sources. The flux density of the compact calibrator OQ~208 with respect to 3C~286 was derived using the inner three short-baseline stations, and the flux density of J0834$-$0417 was derived with respect to OQ~208 and using all the stations. Fourth, we applied the self-calibration solutions and transferred the solutions from J0834$-$0417 to NGC~2617.

The deconvolution was done in \textsc{difmap}. The phase calibrator J0834$-$0417 had a point source structure at 1.5 and 5.5~GHz. It had total flux densities $110 \pm 3$~mJy at 1.5~GHz and $73 \pm 6$~mJy at 5.5~GHz. The imaging results of NGC~2617 are reported in Table~\ref{tab:flux}. We used purely natural grid weighting at C band, and uniform grid weighting at L band.   

\subsection{VLA and VLBA archival data}
We also analysed some historical VLA data to search for the flux density variability. These experiments were observed on 1993 April 26, 1993 July 06 and 1995 July 10. We downloaded these publicly available data from the NRAO data archive\footnote{\url{https://archive.nrao.edu/archive/advquery.jsp}} and reduced them step by step in the standard way recommended by the online \textsc{aips} cookbook\footnote{\url{http://www.aips.nrao.edu/cook.html}} in Appendix A. The final VLA imaging results are listed in Table~\ref{tab:flux}. 

To make some comparisons with our EVN results, we also re-visited the VLBA archival data published by \citep{Jencson2013}. The data reduction followed the new calibration strategy suggested by the \textsc{aips} cookbook in Appendix C. However, fringe fitting followed the EVN calibration strategy to gain more robust solutions in particular for long-baseline stations. The final VLBA imaging results are listed in Table~\ref{tab:flux}.

\section{Multi-resolution imaging results}
\subsection{Parsec-scale radio structure in NGC~2617}
\label{sec:results}
\begin{table*}
\caption{Summary of the circular-Gaussian model fitting results of the visibility data in \textsc{difmap} and the related physical parameters. Columns give (1) component name reported in Fig.~\ref{fig:ngc2617}, (2) observing frequency, (3) peak brightness,  (4) integrated flux density, (5--6) relative offsets in Right Ascension and Declination with respect to the peak component C (RA = 08$^{\rm h}$35$^{\rm m}$38$\fs$798152, Dec = $-$04$\degr$05$\arcmin$17$\farcs$89978, J2000, see also Table~\ref{tab:pos} and Fig.~\ref{fig:core}), (7) full width at half maximum, i.e. de-convolved size, (8) brightness temperature and (9) radio luminosity. The errors in columns (3--7) are the formal errors derived by the program \texttt{modelfit}. The errors in columns (8--9) are the total errors including the systematic errors (ten percent) of flux densities.}
\label{tab:fit}
\centering
\begin{tabular}{cccrrrccc}
\hline
Name & $\nu_{\rm obs}$ &$S_{\rm pk}$       & $S_{\rm int}$ & $\Delta$RA   &  $\Delta$Dec  &  $\theta_{\rm size}$ 
                                                                                                              & $T_{\rm b}$ & $L_{\rm R}$ \\
     & (GHz)          & (mJy beam$^{-1}$)  &  (mJy)        & (mas)          &  (mas)          & (mas)         & (K)         & (erg\,s$^{-1}$) \\
\hline
C    & 4.996          & $1.34\pm0.02$      & $1.21\pm0.02$ & $0.00\pm0.04$  & $0.00\pm0.02$  & $0.35\pm0.06$  & $(4.7\pm1.7)\times10^8$ & $(3.0\pm0.3)\times10^{37}$    \\  
C    & 1.662          & $0.90\pm0.04$      & $0.93\pm0.08$ & $ 0.00\pm0.62$ & $ 0.00\pm0.19$ & $1.56\pm0.35$  & $(1.7\pm0.8)\times10^8$ & $(7.6\pm1.0)\times10^{36}$    \\
J    & 1.662          & $0.40\pm0.04$      & $0.52\pm0.08$ & $-3.10\pm1.00$ & $+4.42\pm0.88$ & $3.46\pm1.40$  & $(1.9\pm1.6)\times10^7$ & $(4.2\pm0.8)\times10^{36}$    \\
\hline
\end{tabular}
\end{table*}

\begin{figure*}
\includegraphics[width=\textwidth]{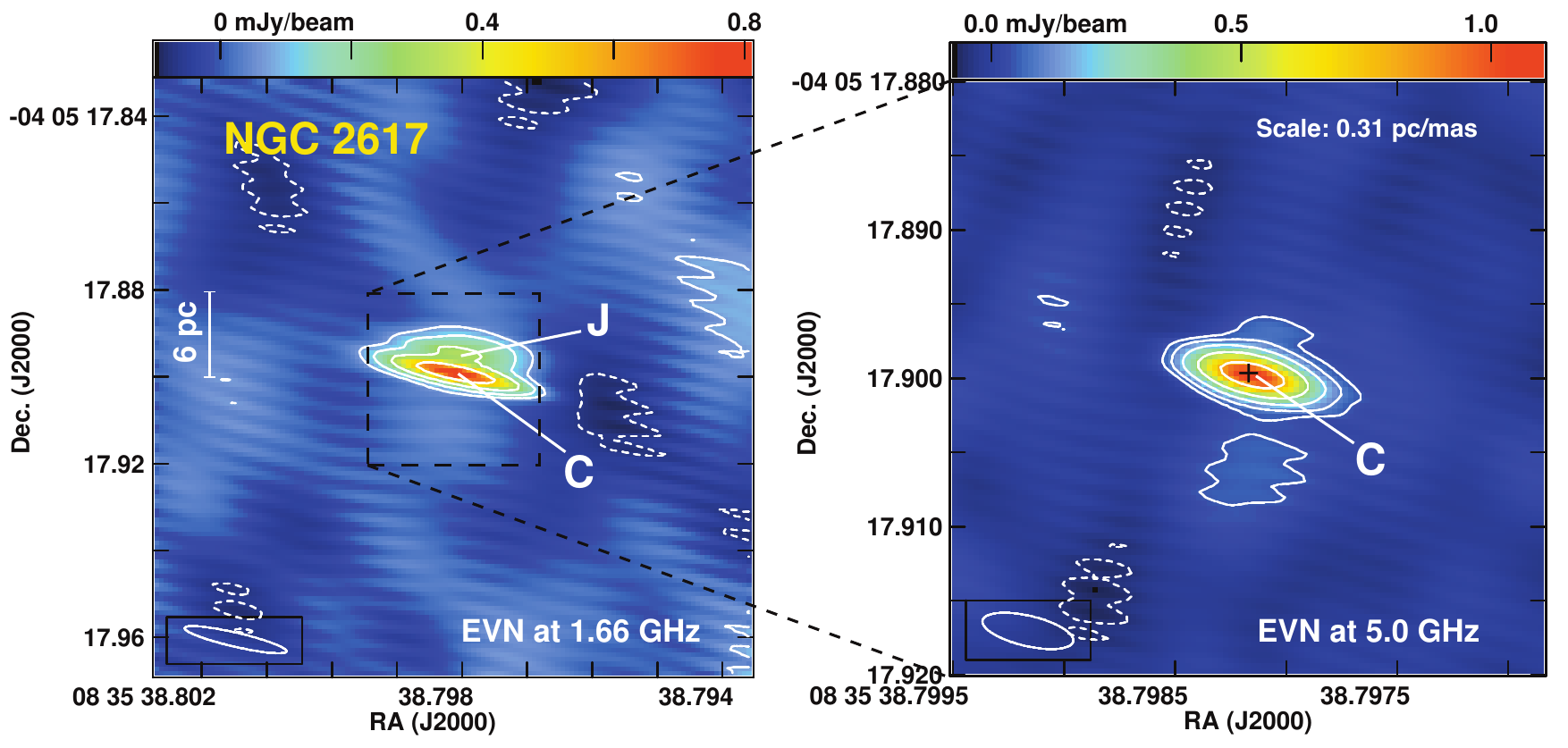}  \\
\caption{
The faint core-jet structure found in the nearby changing-look Seyfert galaxy NGC~2617. \textbf{Left}: The EVN image at 1.66~GHz. The restoring beam FWHM is 24.0~$\times$~3.2~mas$^{2}$ at PA~=~$+77\fdg0$. The contours start from 0.081~mJy\,beam$^{-1}$ (2.5$\sigma$) and increase by factors of $-$1, 1, 2, 4 and 8. The peak brightness is 0.80~mJy\,beam$^{-1}$. \textbf{Right}: The EVN image at 5.0~GHz. The black cross denotes the optical centroid reported by the \textit{Gaia} DR2 (second data release). The beam is 6.3~$\times$~2.0~mas$^{2}$ at PA~=~76$\fdg$9. The contours start from 0.049~mJy\,beam$^{-1}$ (2.5$\sigma$) and increase by factors of $-$1, 1, 2, 4 and 8. The peak brightness is 1.10~mJy\,beam$^{-1}$.}
\label{fig:ngc2617}
\end{figure*}

Fig.~\ref{fig:ngc2617} displays the \textsc{clean} maps of NGC~2617 made with purely natural grid weighting. In the low-resolution intensity image observed at 1.7~GHz on 2013 September 18, NGC~2617 displays a resolved structure, which can be decomposed into two components, C and J.  The secondary component J was not detected in the 1.6-GHz VLBA maps presented by \citet{Jencson2013} and by us in Appendix Fig.~\ref{fig:vlba}, because they have about three times lower sensitivity and  about four times poorer resolution in the north-south direction than the EVN map (c.f. Table~\ref{tab:flux}). In the high-resolution image observed at 5.0~GHz on 2013 October 9, only the peak component C is clearly detected. Based on the \texttt{clean} components, NGC~2617 has total flux densities $1.47 \pm 0.15$~mJy at 1.7~GHz and $1.24 \pm 0.13$~mJy at 5.0~GHz. The uncertainties of the flux densities have included the empirical systematic errors, about ten percent of the flux densities. Within the 3$\sigma$ error bars, the total flux density measurements agree with the VLBA results. If uniform grid weighting is used, then there is a hint of a very faint ($\sim$0.2~mJy\,beam$^{-1}$) and short ($\sim$2~mas) extension towards the north in the EVN map at 5.0~GHz on 2013 October 9. The EVN map is attached in Appendix Fig.~\ref{fig:evn_c_uw}. However, because of its faintness ($\sim5\sigma$) and the very non-optimal ($u$, $v$) coverage of the observations, we cannot fully exclude the possibility of this being a fake structure.

To characterise the components displayed in Fig.~\ref{fig:ngc2617}, we fitted circular Gaussian models to the visibility data in \textsc{difmap} and listed the best-fit parameters in Table~\ref{tab:fit}. The formal uncertainties are derived by the normalisation of the reduced $\chi^2 = 1$. Component J is located $5.4 \pm 1.0$~mas ($1.6 \pm 0.3$~pc) at PA~=~$-35 \pm 12$~deg from the peak component C. The component C shows a relatively compact structure at both 1.7 and 5~GHz, and has a flat radio spectrum with a spectral index of $\alpha = 0.24 \pm 0.15$. Throughout the paper, we define the $\alpha$ to meet the formula $S_{\rm \nu} \propto \nu^{\alpha}$. At 5.0~GHz, the component J is not detected. This is mainly because of its steep radio spectrum and extended structure. When the 1.6-GHz image has a beam equal to that of the 5.0-GHz map, the component J has a weaker peak brightness of 0.22~mJy\,beam$^{-1}$. Using the peak brightness and 3$\sigma$ as the upper limit of its peak brightness at 5.0~GHz, we could provide a constraint of $\alpha \leq -1.2$ for its peak. Similar faint steep-spectrum components are observed in other radio sources \citep[e.g.][]{Middelberg2004, Bontempi2012, Argo2013}.  

With respect to the phase-referencing calibrator J0834$-$0417, we measured the precise position of the peak component C at 5.0~GHz and reported the position in Table~\ref{tab:pos}. Moreover, the optical \textit{Gaia} DR2 astrometry results are also listed in Table~\ref{tab:pos}. The high-precision EVN position is fully consistent with the optical Gaia position \citep{Gaia2018}. The systematic positional errors of the VLBI differential astrometry are completely dominated by the errors of the phase-referencing calibrator position. Because of the very small separation (16~arcmin) between the target and the calibrator, the systematic positional errors due to the ionospheric and tropospheric propagation effects are expected to be quite small \citep[e.g.,][]{Reid2014, Kirsten2015}. The empirical estimate of the systematic positional error is $\sim$0.03~mas according to the five-epoch phase-referencing observations of a pair of faint calibrators with a comparable separation of 14~arcmin at 5~GHz \citep{Mohan2020}.

\begin{table*}
\caption{List of the optical \textit{Gaia} and the EVN 5-GHz differential astrometry results of NGC~2617.  In the columns of errors, $\sigma_{f}$ and $\sigma_{\rm s}$ represent the formal and systematic uncertainties respectively. }
\label{tab:pos}
\begin{tabular}{ccccccc}
\hline
 Technique          & Right Ascension & $\sigma_{\rm ra}$ ($\sigma_{f}$, $\sigma_{\rm s}$) & Declination &  $\sigma_{\rm dec}$ ($\sigma_{f}$, $\sigma_{\rm s}$) & Major source of $\sigma_{\rm s}$          \\
                    &  (J2000)        &  (mas)  &      (J2000)    &   (mas)  \\  
\hline   
 \textit{Gaia} DR2  & 08$^{\rm h}$35$^{\rm m}$38$\fs$798172   & $\pm0.06\pm0.40$  & $-$04$\degr$05$\arcmin$17$\farcs$89963  & $\pm0.06\pm0.40$  & Extended nucleus \\
 EVN differential astrometry    & 08$^{\rm h}$35$^{\rm m}$38$\fs$798152   & $\pm0.04\pm0.50$  & $-$04$\degr$05$\arcmin$17$\farcs$89978  & $\pm0.02\pm1.20$  & Calibrator J0834$-$0417 \\
\hline
\end{tabular}
\end{table*}

We list brightness temperature and radio luminosity ($L_{\rm R} = \nu L_\nu$) for each component in the last two columns of Table~\ref{tab:fit}. The average brightness temperature $T_{\rm b}$ is estimated \citep[e.g.][]{Condon1982} as
\begin{equation}
T_\mathrm{b} = 1.22\times10^{9}\frac{S_\mathrm{int}}{\nu_\mathrm{obs}^2\theta_\mathrm{size}^2}(1+z),
\label{eq1}
\end{equation}
where $S_\mathrm{int}$ is the integrated flux density in mJy, $\nu_\mathrm{obs}$ is the observing frequency in GHz, $\theta_\mathrm{size}$ is the FWHM in mas, and $z$ is the redshift. The average brightness temperature is $\sim$10$^8$~K. We note that the poorer angular resolution of the EVN observations in the direction east-west might over-estimate $\theta_{\rm size}$ and thus under-estimate $T_{\rm b}$ to some degree.

\subsection{\textit{e}-MERLIN and VLA imaging results}

The \textit{e}-MERLIN observations at 5.5~GHz show that NGC~2617 has an unresolved structure with a size of $\leq$11~mas and a total flux density of $1.5 \pm 0.2$~mJy on 2014 January 11 (c.f. Appendix Fig.~\ref{fig:emerlin}). Compared to the total flux densities observed early by the EVN and the VLBA at 5~GHz, we do not detect significant variability on timescale of months.

In the \textit{e}-MERLIN image at 1.5~GHz, Appendix Fig~\ref{fig:emerlin}, NGC~2617 also show a compact feature. With uniform grid weighting, the compact feature has a total flux density of $2.0\pm0.2$~mJy. This is slightly higher than the flux density found in the high-resolution VLBI observations, mainly because its large beam collects some extended nuclear emission. There might also exist some very diffuse emission with a total flux density of $\sim$2~mJy mainly in the north-south direction.         

\begin{figure}
\includegraphics[width=\columnwidth]{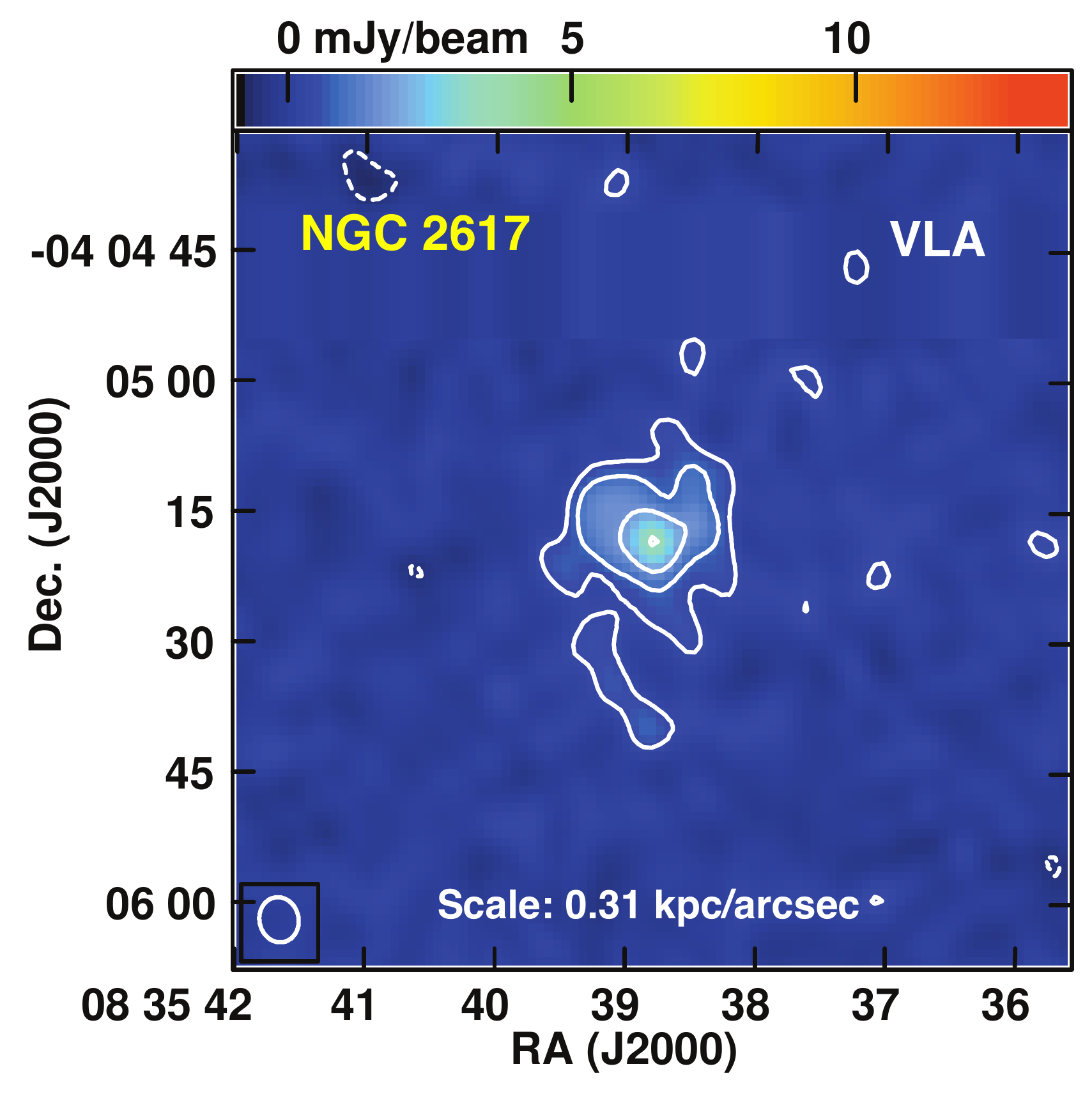}  \\
\caption{
The 1.4-GHz image of the host galaxy of NGC 2617 derived from the VLA observation on 1993 April 26. The contours start from 0.44 mJy~beam$^{-1}$ (2.5$\sigma$) and increase by factors of $-$1, 1, 2, 4 and 8. Other map parameters are presented in Table~\ref{tab:flux}.}
\label{fig:vla}
\end{figure}

The low-resolution VLA observations at 1.4~GHz reveal more diffuse radio emission with a size of 30~arcsec \citep{Condon1982}. Fig.~\ref{fig:vla} shows the diffuse structure of the central part of the host galaxy observed on 1993 April 26. In our highest angular resolution 1.4~GHz VLA data (A-configuration observations on 1995 July 10), only the central nuclear component was detected. 

\section{Discussion}
\label{sec:discussion}
\subsection{Parsec-scale core-jet structure}

\begin{figure}
\includegraphics[width=\columnwidth]{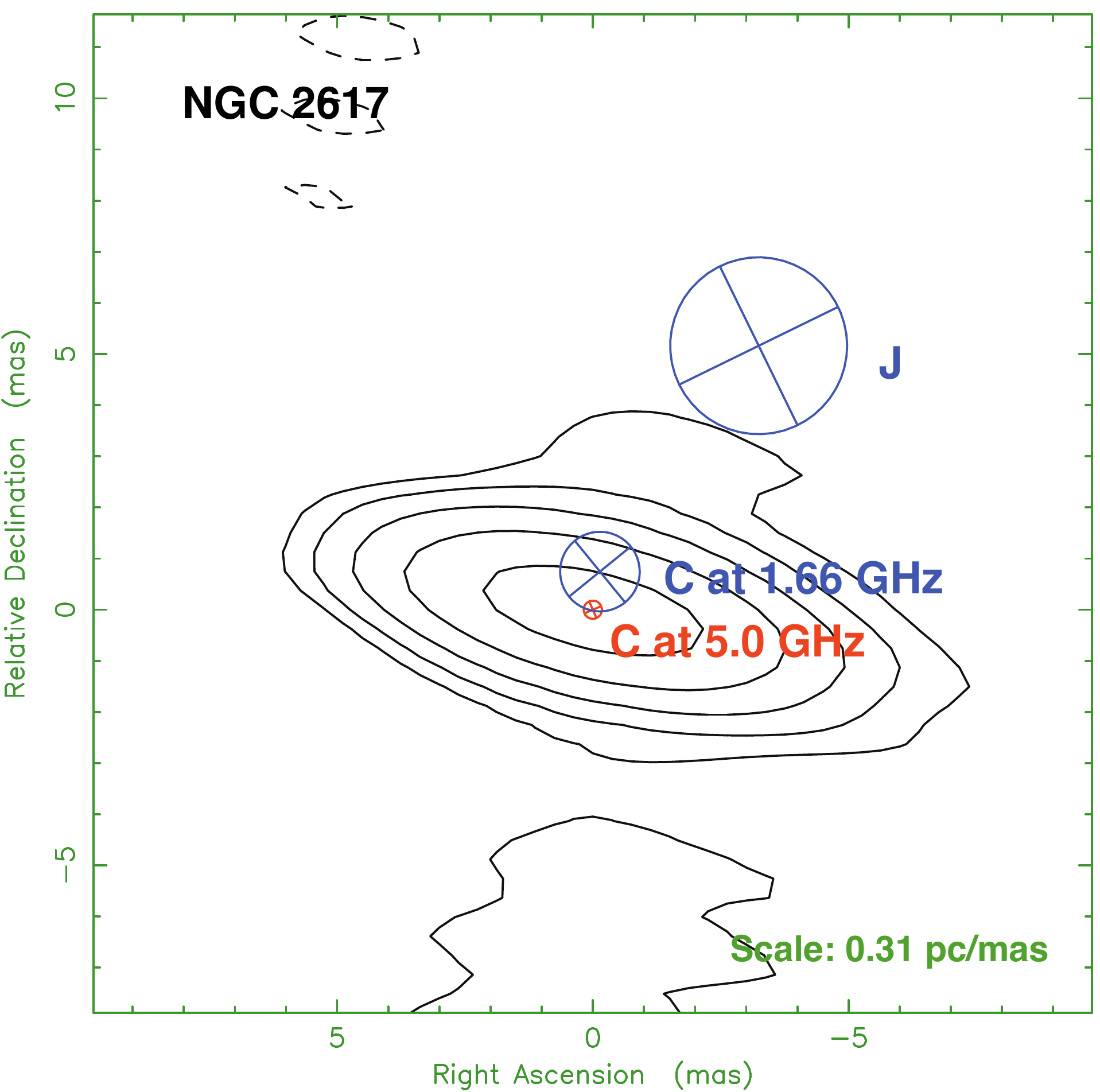}  \\
\caption{
The frequency-dependent positional shift of the radio core in NGC 2617. These circles and their diameters represent the positions and the sizes of the best-fit circular Gaussian models listed in Table~\ref{tab:fit}. }
\label{fig:core}
\end{figure}

The pc-scale radio structure in NGC~2617 can be naturally interpreted as a faint jet powered by the central SMBH. The peak component C can be unambiguously identified as the jet base. First, it has a compact structure and a relatively flat radio spectrum (c.f. Section~\ref{sec:results}). Second, the peak component C at 5.0~GHz is very close ($\sim$0.18~mas) to the centroid of the optical brightness distribution reported by \textit{Gaia} DR2. Finally, there exists a significant frequency-dependent positional shift along the jet direction from 5.0 to 1.7~GHz. Fig.~\ref{fig:core} shows the positional offset: $\Delta$RA~= $-0.14 \pm 0.62$~mas and $\Delta$Dec~= $0.75 \pm 0.19$~mas ($0.23 \pm 0.06$~pc). Because the radio core of the phase-referencing calibrator has a jet direction along the RA, almost perpendicular to the jet direction of NGC~2617, the offset $\Delta$Dec is mainly contributed by the radio core of NGC~2617. This offset can not result from an apparent superluminal motion at a speed of $\sim$12$c$ between the two epochs (separation: 22~d) because of its unusual moving direction (opposite to component J) and no evidence of an extremely relativistic jet. The very short interval 22~d does not allow us to explain the large offset ($\sim$0.7 light year) as the irregular core motion, i.e. jitter \citep[e.g.][]{Yang2016,Rioja2020} because of variable opacity.   

Assuming a frequency dependence of $\nu^{-1}$, the core shift between 1.6 and 5.0~GHz is consistent with the typical value of 0.5~mas between 2.3 and 8.4~GHz observed in bright radio jets \citep[e.g.][]{Plavin2019, Pashchenko2020}. The small angular difference $\Delta \theta$ of the radio core between two observing frequencies in a synchrotron self-absorbed jet \citep{Blandford1979} also scales with its radio luminosity $L_{\rm R}$ and luminosity distance $D_{\rm L}$ as $\Delta \theta \propto D_{\rm L}^{-1} L_{\rm R}^{2/3}$ \citep[Eq.~3,][]{Kovalev2008}. The dependence on the radio luminosity was demonstrated during the outburst of a microquasar \citep{Paragi2013}. Compared to bright ($\ga$1~Jy) one-sided jets with core shifts \citep[e.g.][]{Kovalev2008, Plavin2019}, the jet of NGC~2617 is about three orders of magnitude fainter. However, it has a very small distance ($\sim$60~Mpc) allowing small core shifts to be studied. Within the population of mJy radio jets, these observations are among the first to detect a significant core shift. Other examples include the low-luminosity jets of M~81 which has a radio luminosity comparable to NGC~2617 and also shows a significant core shift \citep{Marti-Vidal2011}.   

Component J can be naturally interpreted as a non-thermal jet component. Its radio luminosity is below the maximum luminosity, $L_{\rm R}$ $\sim$10$^{38.7}$~erg\,s$^{-1}$, observed in the young supernovae \citep{Weiler2002, Varenius2019}. However, it cannot be interpreted as a young supernova because there was no indication of a rapid fading phase \citep[e.g.][]{Varenius2019} during our multi-epoch observations at L band. Furthermore, there was no optical supernova reported in the nucleus of the nearby face-on galaxy NGC~2617.

A core-jet structure is often seen in radio-loud Seyfert galaxies \citep[e.g.][]{Ulvestad2003, Middelberg2004}. The origin of radio emission from radio-quiet AGN might be star formation, an AGN-driven wind, free-free emission from photoionized gas, low-power jets, and the innermost accretion disk coronal activity \citep{Panessa2013, Panessa2019}. Faint radio cores on parsec scales have been detected by VLBI observations in many galaxies \citep[e.g.][]{Giroletti2009, Bontempi2012, Park2017, Gabanyi2018} and quasars \citep[e.g.][]{Yang2012, Yang2021pds}. In the nearby changing-look Seyfert galaxy NGC~2617, the detections of the faint core-jet structure and the core shift provide more direct evidence of the collimated jet activity of the central SMBH. Moreover, compared to 280 nearby galaxies at distances $<$100~Mpc surveyed at high angular resolution with {\emph e-}MERLIN \citep[][]{Baldi2018, Baldi2021}, the radio luminosity of NGC~2617 is a typical value given the distance of $\sim$60~Mpc. 

\subsection{Relation between jet activity, the Seyfert type change and the outburst of 2013}
In NGC~2617, there is no evidence of potential jet activity associated with the previous dramatic Seyfert type change and the large `inside-out' outburst of 2013 Spring. To date, there are only a few reports of jets found in changing-look AGN, such as NGC~4151 \citep[e.g.][]{Penston1984, Williams2020} and Mrk~590 \citep{Koay2016VLBA, Yang2021}. However, only Mrk~590 displayed a coincident radio outburst \citep{Koay2016VLBA} and might have launched a short-lived jet component \citep{Yang2021} during the forty-year dramatic accretion activity \citep{Denney2014}. 

The northern jet component J cannot result from the multi-band outburst of 2013 Spring reported by \citet{Shappee2014}. If it were ejected during the outburst, an apparent jet speed of $\ga$10~$c$ is expected. This inferred speed is too high for a low-power radio source \citep[e.g.][]{Ulvestad2003, Kunert2010, An2012}. Additionally, no radio variability was detected at 5~GHz in the radio core of NGC~2617 over our three epoch observations between 2013 June 29 and 2014 January 11, and as such there is no evidence for a radio outburst associated with the 2013 Spring multi-band outburst. 

The lack of relativistic ejecta in NGC~2617 might be because it failed to transfer to the high accretion rate state during the outburst.  \citet{Ruan2019B} reported that NGC~2617 had a peak Eddington ratio of $\frac{L_{\rm Bol}}{L_{\rm Edd}} < 10^{-2}$. The critical value to discriminate high and low states is likely $\sim$10$^{-2}$ \citep{Ruan2019A}. This is also consistent with the value observed in stellar-mass black hole X-ray binaries \citep[e.g.][]{McClintock2006}. If the unified X-ray outburst model in black hole X-ray binaries \citep{Fender2009} is applicable to extra-galactic AGN \citep{Marscher2002, Yang2021}, we would expect to see a compact radio core and no episodic relativistic ejection event at the low-accretion rate state. These expectations are fully consistent with the VLBI results.  

At the low accretion rate state, there exists a correlation \citep[e.g.][]{Merloni2003} between the radio core luminosity at 5.0~GHz, the X-ray luminosity ($L_{\rm X}$) in the 2--10~keV band, and the BH mass ($M_{\rm bh}$):
\begin{equation}
\log L_{\rm R} =(0.60^{+0.11}_{-0.11}) \log L_{\rm X} + (0.78^{+0.11}_{-0.09}) \log M_{\rm bh} + 7.33^{+4.05}_{-4.07} 
\label{eq2}
\end{equation}
The central black hole in NGC~2617 has a mass of $(4 \pm 1) \times 10^7$~M$_{\sun}$ \citep{Shappee2014}. The X-ray luminosity was $L_{\rm X} = 10^{43.25 \pm 0.15}$~erg\,s$^{-1}$ in the 2--10~keV band on 2013 April 27 and May 24 \citep{Hernndez-Garc2017}. According to Equation~\ref{eq2}, we would expect $L_{\rm R}=10^{39.2 \pm 1.0}$ erg\,s$^{-1}$. This estimate could be two order of magnitude higher while still in the acceptable range in view of the large scatter of the correlation and significant X-ray variability \citep[e.g. about one order of magnitude,][]{Shappee2014}. 

The Seyfert type change is estimated to have occurred between 2010 October and 2012 February \citep{Oknyansky2017}. It is not clear whether the component J was associated with this type change event. NGC~2617 was not detected ($\leq0.3$~mJy beam$^{-1}$) at 8.4~GHz on 1995 July 10 (Table~\ref{tab:flux}). This 8.4\,GHz non-detection may indicate that either the radio core has a very steep spectrum above 5\,GHz, or that the core has undergone a change in flux density between 1995 and 2013 which may have coincided with the reported Seyfert type change event. If the component J is associated with this state change event, and it keeps moving out and fading for the rest of its life, future VLBI observations would be able to detect its proper motion and variability, and then determine its birth time.

\subsection{Short-lived radio sources and optically selected changing-look AGN}
Most low-radio-power sources are probably short-lived jets on timescales of $\la$10$^5$~yr \citep[e.g.][]{AnBaan2012, Woowska2017}. Because of low accretion rates and short active phases, their jets are poorly developed and sometimes disrupted, and thus show relatively compact radio morphology. This is likely the reason why NGC~2617 has a compact core-jet structure. Recently, there are also extremely short-lived radio sources on timescales down to a few years found in the radio surveys \citep[e.g. ][]{Mooley2016, Kunert2020, Nyland2020}. It is not clear whether short-lived radio sources are generally associated with variable accretion activity of changing-look AGN. However, since intensive accretion events may launch relativistic jets at speeds $\ga$0.1~$c$ \citep[e.g.][]{Marscher2002, Fender2009, Argo2015}, changing-look AGN resulting from extreme accretion activity might be associated with some very short-lived radio sources \citep{Woowska2017, Yang2021}. 

Optically changing-look AGN represent unstable accretion systems. Because of variable accretions, their jets might have shorter lives than normal AGN jets. If the hypothesis is true in the local $z < 1$ Universe, this would cause a low detection rate of optically changing-look AGN at radio. \citet{Yang2018} presented a big sample of 26 optically changing-look AGN including five previously known sources at $0.08 < z < 0.58$ in the northern sky. We searched for their radio counterparts at 1.4~GHz in the NVSS catalogue \citep{Condon1998NVSS}. With an image sensitivity of 3$\sigma=1$~mJy\,beam$^{-1}$ in the survey NVSS, there are only three detections: WISEA J101152.99$+$544206.3, WISEA J110455.17$+$011856.6 and FBQS J115227.5$+$320959. We also searched for their radio counterparts in the Faint Images of the Radio Sky at Twenty centimetre \citep[FIRST,][]{Becker1995} and no further sources from the sample of \citet{Yang2018} were detected despite FIRST high sensitivity. In another study including six changing-look quasars at $z \leq 0.4$ reported by \citep{Ruan2019A}, none were found in the radio surveys NVSS and FIRST. In the sample of six nearby ($z \leq 0.17$) changing-look LINER (low-ionization nuclear emission-line region) galaxies \citep{Frederick2019} which transferred from LINER galaxies to Seyfert 1 galaxies or quasars, there is only one radio detection (ZTF18aasszwr). According to these three samples, the radio detection rate of changing-look AGN is about 11 per cent (4/36). If we add five well-studied nearby changing-look AGN detected in FIRST: Mrk~590 \citep[e.g.][]{Koay2016VLBA}, Mrk~1018 \citep[e.g.][]{Noda2018}, 1ES~1927$+$654 \cite[e.g.][]{Ricci2020}, NGC~4151 \citep{Williams2020} and NGC~2617, the detection rate is doubled (9/41). \citet{MacLeod2019} selected 17 changing-look quasars at $z < 0.83$ and without radio counterparts. Even withstanding the inherent selection biases of these combined studies it is clear that the overall radio detection rate (9/58 or $\sim$16percent) of change-looking sources is low.

Using the VLA FIRST survey radio catalogue of April 2003, \citet{Wadadekar2004} searched for radio emission from $\sim$2840 AGN \citep[from the 10th edition of the AGN catalogue,][]{Veron-Cetty2001} and got a detection rate of 27 per cent. The \textit{e}-MERLIN surveys of optically selected local active (LINER and Seyfert) and inactive (H~II galaxies and absorption line galaxies) galaxies gives a radio detection rate of 40 per cent for the central active SMBH \citep{Baldi2021}. Compared to the detection rates of these two larger and more complete AGN samples, the detection rate of the optically changing-look AGN is low. Moreover, we caution that flux densities from the low-resolution surveys NVSS and FIRST can be contaminated to a certain degree by star formation in host galaxies \citep[e.g.][]{Deller2014, Ruiz2017}. The above comparison of the detection rate assumes that both changing-look AGN and general AGN have very similar contamination fractions. If their contamination fractions are significant different, this would give us a systematic bias for the comparison. 

Our current knowledge of the radio properties of the change-look AGN population is mainly limited by small and incomplete sample statistics. However, this apparent low radio detection rate supports the hypothesis that change-looking AGN may be short-lived radio sources. Moreover, similar to normal AGN, the majority of changing-look AGN at $z < 0.83$ are faint sub-mJy sources at radio.

\section{Conclusions}
\label{sec:conclusions}
With the EVN at 1.7 and 5.0~GHz and the \textit{e}-MERLIN at 1.5 and 5.5~GHz, we performed high-resolution radio observations of the nearby changing-look Seyfert galaxy NGC~2617 to probe its potential jet activity. We found that there exists a flat-spectrum compact radio core and a steep-spectrum jet component launched by the SMBH of NGC~2617. The one-sided jet extends towards the north up to two parsecs at 1.7~GHz and shows a core shift of $0.8 \pm 0.2$~mas between 1.7 and 5.0~GHz. We also re-visited the VLA and VLBA archival data. Our observations show that the radio core had no significant radio variability after the multi-band `inside-out' outburst of 2013 Spring. The northern jet might result from a much early outburst. Furthermore, we searched for radio counterparts of some changing-look AGN samples at $z \leq 0.83$ in the NVSS and FIRST catalogues, and noticed that changing-look AGN are dominated by sub-mJy radio sources and might be less active than normal AGN at radio.

\section*{Acknowledgements}
TA and XW are supported by the National Key R\&D Programme of China (2018YFA0404603) and the Chinese Academy of Sciences(CAS, 114231KYSB20170003).
LF acknowledges the support from the National Natural Science Foundation of China (NSFC, grant Nos. 11822303, 11773020).
The European VLBI Network (EVN) is a joint facility of independent European, African, Asian, and North American radio astronomy institutes. Scientific results from data presented in this publication are derived from the following EVN project code(s): RY005 and EY021. 
e-VLBI research infrastructure in Europe is supported by the European Union’s Seventh Framework Programme (FP7/2007-2013) under grant agreement number RI-261525 NEXPReS.
\textit{e}-MERLIN is a National Facility operated by the University of Manchester at Jodrell Bank Observatory on behalf of STFC.
The research leading to these results has received funding from the European Commission Seventh Framework Programme (FP/2007-2013) under grant agreement No. 283393 (RadioNet3).
This work has made use of data from the European Space Agency (ESA) mission {\it Gaia} (\url{https://www.cosmos.esa.int/gaia}), processed by the {\it Gaia} Data Processing and Analysis Consortium (DPAC, \url{https://www.cosmos.esa.int/web/gaia/dpac/consortium}). Funding for the DPAC has been provided by national institutions, in particular the institutions participating in the {\it Gaia} Multilateral Agreement.
This research has made use of the NASA/IPAC Extragalactic Database (NED), which is operated by the Jet Propulsion Laboratory, California Institute of Technology, under contract with the National Aeronautics and Space Administration.
This research has made use of NASA’s Astrophysics Data System Bibliographic Services.

\section*{Data Availability}
The correlation data underlying this article are available in the EVN data archive (\url{http://www.jive.nl/select-experiment}). The calibrated visibility data underlying this article will be shared on reasonable request to the corresponding author.
 



\bibliographystyle{mnras}
\bibliography{NGC2617} 

\begin{thebibliography}{}
\makeatletter
\relax
\def\mn@urlcharsother{\let\do\@makeother \do\$\do\&\do\#\do\^\do\_\do\%\do\~}
\def\mn@doi{\begingroup\mn@urlcharsother \@ifnextchar [ {\mn@doi@}
  {\mn@doi@[]}}
\def\mn@doi@[#1]#2{\def\@tempa{#1}\ifx\@tempa\@empty \href
  {http://dx.doi.org/#2} {doi:#2}\else \href {http://dx.doi.org/#2} {#1}\fi
  \endgroup}
\def\mn@eprint#1#2{\mn@eprint@#1:#2::\@nil}
\def\mn@eprint@arXiv#1{\href {http://arxiv.org/abs/#1} {{\tt arXiv:#1}}}
\def\mn@eprint@dblp#1{\href {http://dblp.uni-trier.de/rec/bibtex/#1.xml}
  {dblp:#1}}
\def\mn@eprint@#1:#2:#3:#4\@nil{\def\@tempa {#1}\def\@tempb {#2}\def\@tempc
  {#3}\ifx \@tempc \@empty \let \@tempc \@tempb \let \@tempb \@tempa \fi \ifx
  \@tempb \@empty \def\@tempb {arXiv}\fi \@ifundefined
  {mn@eprint@\@tempb}{\@tempb:\@tempc}{\expandafter \expandafter \csname
  mn@eprint@\@tempb\endcsname \expandafter{\@tempc}}}

\bibitem[\protect\citeauthoryear{{An} \& {Baan}}{{An} \&
  {Baan}}{2012}]{AnBaan2012}
{An} T.,  {Baan} W.~A.,  2012, \mn@doi [\apj] {10.1088/0004-637X/760/1/77},
  \href {https://ui.adsabs.harvard.edu/abs/2012ApJ...760...77A} {760, 77}

\bibitem[\protect\citeauthoryear{{An} et~al.,}{{An} et~al.}{2012}]{An2012}
{An} T.,  et~al., 2012, \mn@doi [\apjs] {10.1088/0067-0049/198/1/5}, \href
  {https://ui.adsabs.harvard.edu/abs/2012ApJS..198....5A} {198, 5}

\bibitem[\protect\citeauthoryear{{Argo}, {Paragi}, {Rottgering}, {Klockner},
  {Miley}  \& {Mahmud}}{{Argo} et~al.}{2013}]{Argo2013}
{Argo} M.~K.,  {Paragi} Z.,  {Rottgering} H.,  {Klockner} H.~R.,  {Miley} G.,
  {Mahmud} M.,  2013, \mn@doi [\mnras] {10.1093/mnrasl/slt008}, \href
  {https://ui.adsabs.harvard.edu/abs/2013MNRAS.431L..58A} {431, L58}

\bibitem[\protect\citeauthoryear{{Argo}, {van Bemmel}, {Connolly}  \&
  {Beswick}}{{Argo} et~al.}{2015}]{Argo2015}
{Argo} M.~K.,  {van Bemmel} I.~M.,  {Connolly} S.~D.,   {Beswick} R.~J.,  2015,
  \mn@doi [\mnras] {10.1093/mnras/stv1348}, \href
  {https://ui.adsabs.harvard.edu/abs/2015MNRAS.452.1081A} {452, 1081}

\bibitem[\protect\citeauthoryear{{Baldi} et~al.,}{{Baldi}
  et~al.}{2018}]{Baldi2018}
{Baldi} R.~D.,  et~al., 2018, \mn@doi [\mnras] {10.1093/mnras/sty342}, \href
  {https://ui.adsabs.harvard.edu/abs/2018MNRAS.476.3478B} {476, 3478}

\bibitem[\protect\citeauthoryear{{Baldi} et~al.,}{{Baldi}
  et~al.}{2021}]{Baldi2021}
{Baldi} R.~D.,  et~al., 2021, \mn@doi [\mnras] {10.1093/mnras/staa3519}, \href
  {https://ui.adsabs.harvard.edu/abs/2021MNRAS.500.4749B} {500, 4749}

\bibitem[\protect\citeauthoryear{{Becker}, {White}  \& {Helfand}}{{Becker}
  et~al.}{1995}]{Becker1995}
{Becker} R.~H.,  {White} R.~L.,   {Helfand} D.~J.,  1995, \mn@doi [\apj]
  {10.1086/176166}, \href
  {https://ui.adsabs.harvard.edu/abs/1995ApJ...450..559B} {450, 559}

\bibitem[\protect\citeauthoryear{{Blandford} \& {K{\"o}nigl}}{{Blandford} \&
  {K{\"o}nigl}}{1979}]{Blandford1979}
{Blandford} R.~D.,  {K{\"o}nigl} A.,  1979, \mn@doi [\apj] {10.1086/157262},
  \href {https://ui.adsabs.harvard.edu/abs/1979ApJ...232...34B} {232, 34}

\bibitem[\protect\citeauthoryear{{Blandford}, {Meier}  \&
  {Readhead}}{{Blandford} et~al.}{2019}]{Blandford2019}
{Blandford} R.,  {Meier} D.,   {Readhead} A.,  2019, \mn@doi [\araa]
  {10.1146/annurev-astro-081817-051948}, \href
  {https://ui.adsabs.harvard.edu/abs/2019ARA&A..57..467B} {57, 467}

\bibitem[\protect\citeauthoryear{{Bontempi}, {Giroletti}, {Panessa}, {Orienti}
  \& {Doi}}{{Bontempi} et~al.}{2012}]{Bontempi2012}
{Bontempi} P.,  {Giroletti} M.,  {Panessa} F.,  {Orienti} M.,   {Doi} A.,
  2012, \mn@doi [\mnras] {10.1111/j.1365-2966.2012.21786.x}, \href
  {https://ui.adsabs.harvard.edu/abs/2012MNRAS.426..588B} {426, 588}

\bibitem[\protect\citeauthoryear{{Condon}, {Condon}, {Gisler}  \&
  {Puschell}}{{Condon} et~al.}{1982}]{Condon1982}
{Condon} J.~J.,  {Condon} M.~A.,  {Gisler} G.,   {Puschell} J.~J.,  1982,
  \mn@doi [\apj] {10.1086/159538}, \href
  {https://ui.adsabs.harvard.edu/abs/1982ApJ...252..102C} {252, 102}

\bibitem[\protect\citeauthoryear{{Condon}, {Cotton}, {Greisen}, {Yin},
  {Perley}, {Taylor}  \& {Broderick}}{{Condon} et~al.}{1998}]{Condon1998NVSS}
{Condon} J.~J.,  {Cotton} W.~D.,  {Greisen} E.~W.,  {Yin} Q.~F.,  {Perley}
  R.~A.,  {Taylor} G.~B.,   {Broderick} J.~J.,  1998, \mn@doi [\aj]
  {10.1086/300337}, \href
  {https://ui.adsabs.harvard.edu/abs/1998AJ....115.1693C} {115, 1693}

\bibitem[\protect\citeauthoryear{{Deller} \& {Middelberg}}{{Deller} \&
  {Middelberg}}{2014}]{Deller2014}
{Deller} A.~T.,  {Middelberg} E.,  2014, \mn@doi [\aj]
  {10.1088/0004-6256/147/1/14}, \href
  {https://ui.adsabs.harvard.edu/abs/2014AJ....147...14D} {147, 14}

\bibitem[\protect\citeauthoryear{{Denney} et~al.,}{{Denney}
  et~al.}{2014}]{Denney2014}
{Denney} K.~D.,  et~al., 2014, \mn@doi [\apj] {10.1088/0004-637X/796/2/134},
  \href {https://ui.adsabs.harvard.edu/abs/2014ApJ...796..134D} {796, 134}

\bibitem[\protect\citeauthoryear{{Doyle} et~al.,}{{Doyle}
  et~al.}{2005}]{Doyle2005}
{Doyle} M.~T.,  et~al., 2005, \mn@doi [\mnras]
  {10.1111/j.1365-2966.2005.09159.x}, \href
  {https://ui.adsabs.harvard.edu/abs/2005MNRAS.361...34D} {361, 34}

\bibitem[\protect\citeauthoryear{{Elitzur}, {Ho}  \& {Trump}}{{Elitzur}
  et~al.}{2014}]{Elitzur2014}
{Elitzur} M.,  {Ho} L.~C.,   {Trump} J.~R.,  2014, \mn@doi [\mnras]
  {10.1093/mnras/stt2445}, \href
  {https://ui.adsabs.harvard.edu/abs/2014MNRAS.438.3340E} {438, 3340}

\bibitem[\protect\citeauthoryear{{Fender}, {Homan}  \& {Belloni}}{{Fender}
  et~al.}{2009}]{Fender2009}
{Fender} R.~P.,  {Homan} J.,   {Belloni} T.~M.,  2009, \mn@doi [\mnras]
  {10.1111/j.1365-2966.2009.14841.x}, \href
  {https://ui.adsabs.harvard.edu/abs/2009MNRAS.396.1370F} {396, 1370}

\bibitem[\protect\citeauthoryear{{Frederick} et~al.,}{{Frederick}
  et~al.}{2019}]{Frederick2019}
{Frederick} S.,  et~al., 2019, \mn@doi [\apj] {10.3847/1538-4357/ab3a38}, \href
  {https://ui.adsabs.harvard.edu/abs/2019ApJ...883...31F} {883, 31}

\bibitem[\protect\citeauthoryear{{Gabanyi}, {Frey}, {Paragi}  \&
  {An}}{{Gabanyi} et~al.}{2014}]{Gabanyi2014}
{Gabanyi} K.~E.,  {Frey} S.,  {Paragi} Z.,   {An} T.,  2014, in 40th COSPAR
  Scientific Assembly. pp E1.19--36--14

\bibitem[\protect\citeauthoryear{{Gab{\'a}nyi}, {Frey}, {Paragi},
  {J{\"a}rvel{\"a}}, {Morokuma}, {An}, {Tanaka}  \& {Tar}}{{Gab{\'a}nyi}
  et~al.}{2018}]{Gabanyi2018}
{Gab{\'a}nyi} K.~{\'E}.,  {Frey} S.,  {Paragi} Z.,  {J{\"a}rvel{\"a}} E.,
  {Morokuma} T.,  {An} T.,  {Tanaka} M.,   {Tar} I.,  2018, \mn@doi [\mnras]
  {10.1093/mnras/stx2449}, \href
  {https://ui.adsabs.harvard.edu/abs/2018MNRAS.473.1554G} {473, 1554}

\bibitem[\protect\citeauthoryear{{Gaia Collaboration} et~al.,}{{Gaia
  Collaboration} et~al.}{2018}]{Gaia2018}
{Gaia Collaboration} et~al., 2018, \mn@doi [\aap]
  {10.1051/0004-6361/201833051}, \href
  {https://ui.adsabs.harvard.edu/abs/2018A&A...616A...1G} {616, A1}

\bibitem[\protect\citeauthoryear{{Giroletti} \& {Panessa}}{{Giroletti} \&
  {Panessa}}{2009}]{Giroletti2009}
{Giroletti} M.,  {Panessa} F.,  2009, \mn@doi [\apjl]
  {10.1088/0004-637X/706/2/L260}, \href
  {https://ui.adsabs.harvard.edu/abs/2009ApJ...706L.260G} {706, L260}

\bibitem[\protect\citeauthoryear{{Greisen}}{{Greisen}}{2003}]{Greisen2003}
{Greisen} E.~W.,  2003, in {Heck} A.,  ed.,  Astrophysics and Space Science
  Library Vol. 285, Information Handling in Astronomy - Historical Vistas.
  Kluwer, p.~109, \mn@doi{10.1007/0-306-48080-8_7}

\bibitem[\protect\citeauthoryear{{Guo} et~al.,}{{Guo} et~al.}{2020}]{Guo2020}
{Guo} H.,  et~al., 2020, \mn@doi [\apj] {10.3847/1538-4357/abc2ce}, \href
  {https://ui.adsabs.harvard.edu/abs/2020ApJ...905...52G} {905, 52}

\bibitem[\protect\citeauthoryear{{Hern{\'a}ndez-Garc{\'\i}a}, {Masegosa},
  {Gonz{\'a}lez-Mart{\'\i}n}, {M{\'a}rquez}, {Guainazzi}  \&
  {Panessa}}{{Hern{\'a}ndez-Garc{\'\i}a} et~al.}{2017}]{Hernndez-Garc2017}
{Hern{\'a}ndez-Garc{\'\i}a} L.,  {Masegosa} J.,  {Gonz{\'a}lez-Mart{\'\i}n} O.,
   {M{\'a}rquez} I.,  {Guainazzi} M.,   {Panessa} F.,  2017, \mn@doi [\aap]
  {10.1051/0004-6361/201730476}, \href
  {https://ui.adsabs.harvard.edu/abs/2017A&A...602A..65H} {602, A65}

\bibitem[\protect\citeauthoryear{{Herrera Ruiz} et~al.,}{{Herrera Ruiz}
  et~al.}{2017}]{Ruiz2017}
{Herrera Ruiz} N.,  et~al., 2017, \mn@doi [\aap] {10.1051/0004-6361/201731163},
  \href {https://ui.adsabs.harvard.edu/abs/2017A&A...607A.132H} {607, A132}

\bibitem[\protect\citeauthoryear{{Hutsem{\'e}kers}, {Ag{\'\i}s Gonz{\'a}lez},
  {Marin}, {Sluse}, {Ramos Almeida}  \& {Acosta Pulido}}{{Hutsem{\'e}kers}
  et~al.}{2019}]{Hutsemkers2019}
{Hutsem{\'e}kers} D.,  {Ag{\'\i}s Gonz{\'a}lez} B.,  {Marin} F.,  {Sluse} D.,
  {Ramos Almeida} C.,   {Acosta Pulido} J.~A.,  2019, \mn@doi [\aap]
  {10.1051/0004-6361/201834633}, \href
  {https://ui.adsabs.harvard.edu/abs/2019A&A...625A..54H} {625, A54}

\bibitem[\protect\citeauthoryear{{Jencson}, {Kundert}, {Mioduszewski}, {Lucy},
  {Kadowaki}  \& {Mellon}}{{Jencson} et~al.}{2013}]{Jencson2013}
{Jencson} J.,  {Kundert} K.,  {Mioduszewski} A.,  {Lucy} A.,  {Kadowaki} J.,
  {Mellon} S.~N.,  2013, The Astronomer's Telegram, \href
  {https://ui.adsabs.harvard.edu/abs/2013ATel.5347....1J} {5347, 1}

\bibitem[\protect\citeauthoryear{{Kaspi}, {Maoz}, {Netzer}, {Peterson},
  {Vestergaard}  \& {Jannuzi}}{{Kaspi} et~al.}{2005}]{Kaspi2005}
{Kaspi} S.,  {Maoz} D.,  {Netzer} H.,  {Peterson} B.~M.,  {Vestergaard} M.,
  {Jannuzi} B.~T.,  2005, \mn@doi [\apj] {10.1086/431275}, \href
  {https://ui.adsabs.harvard.edu/abs/2005ApJ...629...61K} {629, 61}

\bibitem[\protect\citeauthoryear{{Keimpema} et~al.,}{{Keimpema}
  et~al.}{2015}]{Keimpema2015}
{Keimpema} A.,  et~al., 2015, \mn@doi [Experimental Astronomy]
  {10.1007/s10686-015-9446-1}, \href
  {https://ui.adsabs.harvard.edu/abs/2015ExA....39..259K} {39, 259}

\bibitem[\protect\citeauthoryear{{Kettenis}, {van Langevelde}, {Reynolds}  \&
  {Cotton}}{{Kettenis} et~al.}{2006}]{Kettenis2006}
{Kettenis} M.,  {van Langevelde} H.~J.,  {Reynolds} C.,   {Cotton} B.,  2006,
  in {Gabriel} C.,  {Arviset} C.,  {Ponz} D.,   {Enrique} S.,  eds,
  Astronomical Society of the Pacific Conference Series Vol. 351, Astronomical
  Data Analysis Software and Systems XV. Astron. Soc. Pac., p.~497

\bibitem[\protect\citeauthoryear{{Kirsten}, {Vlemmings}, {Campbell}, {Kramer}
  \& {Chatterjee}}{{Kirsten} et~al.}{2015}]{Kirsten2015}
{Kirsten} F.,  {Vlemmings} W.,  {Campbell} R.~M.,  {Kramer} M.,   {Chatterjee}
  S.,  2015, \mn@doi [\aap] {10.1051/0004-6361/201425562}, \href
  {https://ui.adsabs.harvard.edu/abs/2015A&A...577A.111K} {577, A111}

\bibitem[\protect\citeauthoryear{{Koay}, {Vestergaard}, {Bignall}, {Reynolds}
  \& {Peterson}}{{Koay} et~al.}{2016}]{Koay2016VLBA}
{Koay} J.~Y.,  {Vestergaard} M.,  {Bignall} H.~E.,  {Reynolds} C.,   {Peterson}
  B.~M.,  2016, \mn@doi [\mnras] {10.1093/mnras/stw975}, \href
  {https://ui.adsabs.harvard.edu/abs/2016MNRAS.460..304K} {460, 304}

\bibitem[\protect\citeauthoryear{{Kovalev}, {Lobanov}, {Pushkarev}  \&
  {Zensus}}{{Kovalev} et~al.}{2008}]{Kovalev2008}
{Kovalev} Y.~Y.,  {Lobanov} A.~P.,  {Pushkarev} A.~B.,   {Zensus} J.~A.,  2008,
  \mn@doi [\aap] {10.1051/0004-6361:20078679}, \href
  {https://ui.adsabs.harvard.edu/abs/2008A&A...483..759K} {483, 759}

\bibitem[\protect\citeauthoryear{{Kunert-Bajraszewska}, {Gawro{\'n}ski},
  {Labiano}  \& {Siemiginowska}}{{Kunert-Bajraszewska}
  et~al.}{2010}]{Kunert2010}
{Kunert-Bajraszewska} M.,  {Gawro{\'n}ski} M.~P.,  {Labiano} A.,
  {Siemiginowska} A.,  2010, \mn@doi [\mnras]
  {10.1111/j.1365-2966.2010.17271.x}, \href
  {https://ui.adsabs.harvard.edu/abs/2010MNRAS.408.2261K} {408, 2261}

\bibitem[\protect\citeauthoryear{{Kunert-Bajraszewska}, {Wo{\l}owska},
  {Mooley}, {Kharb}  \& {Hallinan}}{{Kunert-Bajraszewska}
  et~al.}{2020}]{Kunert2020}
{Kunert-Bajraszewska} M.,  {Wo{\l}owska} A.,  {Mooley} K.,  {Kharb} P.,
  {Hallinan} G.,  2020, \mn@doi [\apj] {10.3847/1538-4357/ab9598}, \href
  {https://ui.adsabs.harvard.edu/abs/2020ApJ...897..128K} {897, 128}

\bibitem[\protect\citeauthoryear{{LaMassa} et~al.,}{{LaMassa}
  et~al.}{2015}]{LaMassa2015}
{LaMassa} S.~M.,  et~al., 2015, \mn@doi [\apj] {10.1088/0004-637X/800/2/144},
  \href {https://ui.adsabs.harvard.edu/abs/2015ApJ...800..144L} {800, 144}

\bibitem[\protect\citeauthoryear{{MacLeod} et~al.,}{{MacLeod}
  et~al.}{2019}]{MacLeod2019}
{MacLeod} C.~L.,  et~al., 2019, \mn@doi [\apj] {10.3847/1538-4357/ab05e2},
  \href {https://ui.adsabs.harvard.edu/abs/2019ApJ...874....8M} {874, 8}

\bibitem[\protect\citeauthoryear{{Marscher}, {Jorstad}, {G{\'o}mez}, {Aller},
  {Ter{\"a}sranta}, {Lister}  \& {Stirling}}{{Marscher}
  et~al.}{2002}]{Marscher2002}
{Marscher} A.~P.,  {Jorstad} S.~G.,  {G{\'o}mez} J.-L.,  {Aller} M.~F.,
  {Ter{\"a}sranta} H.,  {Lister} M.~L.,   {Stirling} A.~M.,  2002, \mn@doi
  [\nat] {10.1038/nature00772}, \href
  {https://ui.adsabs.harvard.edu/abs/2002Natur.417..625M} {417, 625}

\bibitem[\protect\citeauthoryear{{Mart{\'\i}-Vidal}, {Marcaide}, {Alberdi},
  {P{\'e}rez-Torres}, {Ros}  \& {Guirado}}{{Mart{\'\i}-Vidal}
  et~al.}{2011}]{Marti-Vidal2011}
{Mart{\'\i}-Vidal} I.,  {Marcaide} J.~M.,  {Alberdi} A.,  {P{\'e}rez-Torres}
  M.~A.,  {Ros} E.,   {Guirado} J.~C.,  2011, \mn@doi [\aap]
  {10.1051/0004-6361/201117211}, \href
  {https://ui.adsabs.harvard.edu/abs/2011A&A...533A.111M} {533, A111}

\bibitem[\protect\citeauthoryear{{McClintock} \& {Remillard}}{{McClintock} \&
  {Remillard}}{2006}]{McClintock2006}
{McClintock} J.~E.,  {Remillard} R.~A.,  2006, {Black hole binaries}.
Cambridge Univ. Press, pp 157--213

\bibitem[\protect\citeauthoryear{{McElroy} et~al.,}{{McElroy}
  et~al.}{2016}]{McElroy2016}
{McElroy} R.~E.,  et~al., 2016, \mn@doi [\aap] {10.1051/0004-6361/201629102},
  \href {https://ui.adsabs.harvard.edu/abs/2016A&A...593L...8M} {593, L8}

\bibitem[\protect\citeauthoryear{{Merloni}, {Heinz}  \& {di Matteo}}{{Merloni}
  et~al.}{2003}]{Merloni2003}
{Merloni} A.,  {Heinz} S.,   {di Matteo} T.,  2003, \mn@doi [\mnras]
  {10.1046/j.1365-2966.2003.07017.x}, \href
  {https://ui.adsabs.harvard.edu/abs/2003MNRAS.345.1057M} {345, 1057}

\bibitem[\protect\citeauthoryear{{Middelberg} et~al.,}{{Middelberg}
  et~al.}{2004}]{Middelberg2004}
{Middelberg} E.,  et~al., 2004, \mn@doi [\aap] {10.1051/0004-6361:20040019},
  \href {https://ui.adsabs.harvard.edu/abs/2004A&A...417..925M} {417, 925}

\bibitem[\protect\citeauthoryear{{Mohan}, {An}  \& {Yang}}{{Mohan}
  et~al.}{2020}]{Mohan2020}
{Mohan} P.,  {An} T.,   {Yang} J.,  2020, \mn@doi [\apjl]
  {10.3847/2041-8213/ab64d1}, \href
  {https://ui.adsabs.harvard.edu/abs/2020ApJ...888L..24M} {888, L24}

\bibitem[\protect\citeauthoryear{{Mooley} et~al.,}{{Mooley}
  et~al.}{2016}]{Mooley2016}
{Mooley} K.~P.,  et~al., 2016, \mn@doi [\apj] {10.3847/0004-637X/818/2/105},
  \href {https://ui.adsabs.harvard.edu/abs/2016ApJ...818..105M} {818, 105}

\bibitem[\protect\citeauthoryear{{Moran}, {Halpern}  \& {Helfand}}{{Moran}
  et~al.}{1996}]{Moran1996}
{Moran} E.~C.,  {Halpern} J.~P.,   {Helfand} D.~J.,  1996, \mn@doi [\apjs]
  {10.1086/192341}, \href
  {https://ui.adsabs.harvard.edu/abs/1996ApJS..106..341M} {106, 341}

\bibitem[\protect\citeauthoryear{{Noda} \& {Done}}{{Noda} \&
  {Done}}{2018}]{Noda2018}
{Noda} H.,  {Done} C.,  2018, \mn@doi [\mnras] {10.1093/mnras/sty2032}, \href
  {https://ui.adsabs.harvard.edu/abs/2018MNRAS.480.3898N} {480, 3898}

\bibitem[\protect\citeauthoryear{{Nyland} et~al.,}{{Nyland}
  et~al.}{2020}]{Nyland2020}
{Nyland} K.,  et~al., 2020, \mn@doi [\apj] {10.3847/1538-4357/abc341}, \href
  {https://ui.adsabs.harvard.edu/abs/2020ApJ...905...74N} {905, 74}

\bibitem[\protect\citeauthoryear{{Oknyansky} et~al.,}{{Oknyansky}
  et~al.}{2017}]{Oknyansky2017}
{Oknyansky} V.~L.,  et~al., 2017, \mn@doi [\mnras] {10.1093/mnras/stx149},
  \href {https://ui.adsabs.harvard.edu/abs/2017MNRAS.467.1496O} {467, 1496}

\bibitem[\protect\citeauthoryear{{Panessa} \& {Giroletti}}{{Panessa} \&
  {Giroletti}}{2013}]{Panessa2013}
{Panessa} F.,  {Giroletti} M.,  2013, \mn@doi [\mnras] {10.1093/mnras/stt547},
  \href {https://ui.adsabs.harvard.edu/abs/2013MNRAS.432.1138P} {432, 1138}

\bibitem[\protect\citeauthoryear{{Panessa}, {Baldi}, {Laor}, {Padovani},
  {Behar}  \& {McHardy}}{{Panessa} et~al.}{2019}]{Panessa2019}
{Panessa} F.,  {Baldi} R.~D.,  {Laor} A.,  {Padovani} P.,  {Behar} E.,
  {McHardy} I.,  2019, \mn@doi [Nature Astronomy] {10.1038/s41550-019-0765-4},
  \href {https://ui.adsabs.harvard.edu/abs/2019NatAs...3..387P} {3, 387}

\bibitem[\protect\citeauthoryear{{Paragi} et~al.,}{{Paragi}
  et~al.}{2013}]{Paragi2013}
{Paragi} Z.,  et~al., 2013, \mn@doi [\mnras] {10.1093/mnras/stt545}, \href
  {https://ui.adsabs.harvard.edu/abs/2013MNRAS.432.1319P} {432, 1319}

\bibitem[\protect\citeauthoryear{{Park}, {Yang}, {Oonk}  \& {Paragi}}{{Park}
  et~al.}{2017}]{Park2017}
{Park} S.,  {Yang} J.,  {Oonk} J.~B.~R.,   {Paragi} Z.,  2017, \mn@doi [\mnras]
  {10.1093/mnras/stw3012}, \href
  {https://ui.adsabs.harvard.edu/abs/2017MNRAS.465.3943P} {465, 3943}

\bibitem[\protect\citeauthoryear{{Pashchenko}, {Plavin}, {Kutkin}  \&
  {Kovalev}}{{Pashchenko} et~al.}{2020}]{Pashchenko2020}
{Pashchenko} I.~N.,  {Plavin} A.~V.,  {Kutkin} A.~M.,   {Kovalev} Y.~Y.,  2020,
  \mn@doi [\mnras] {10.1093/mnras/staa3140}, \href
  {https://ui.adsabs.harvard.edu/abs/2020MNRAS.499.4515P} {499, 4515}

\bibitem[\protect\citeauthoryear{{Paturel}, {Theureau}, {Bottinelli},
  {Gouguenheim}, {Coudreau-Durand}, {Hallet}  \& {Petit}}{{Paturel}
  et~al.}{2003}]{Paturel2003}
{Paturel} G.,  {Theureau} G.,  {Bottinelli} L.,  {Gouguenheim} L.,
  {Coudreau-Durand} N.,  {Hallet} N.,   {Petit} C.,  2003, \mn@doi [\aap]
  {10.1051/0004-6361:20031412}, \href
  {https://ui.adsabs.harvard.edu/abs/2003A&A...412...57P} {412, 57}

\bibitem[\protect\citeauthoryear{{Penston} \& {Perez}}{{Penston} \&
  {Perez}}{1984}]{Penston1984}
{Penston} M.~V.,  {Perez} E.,  1984, \mn@doi [\mnras]
  {10.1093/mnras/211.1.33P}, \href
  {https://ui.adsabs.harvard.edu/abs/1984MNRAS.211P..33P} {211, 33P}

\bibitem[\protect\citeauthoryear{{Peterson}, {Korista}  \& {Cota}}{{Peterson}
  et~al.}{1986}]{Peterson1986}
{Peterson} B.~M.,  {Korista} K.~T.,   {Cota} S.~A.,  1986, \baas, \href
  {https://ui.adsabs.harvard.edu/abs/1986BAAS...18.1001P} {18, 1001}

\bibitem[\protect\citeauthoryear{{Petrov}}{{Petrov}}{2021}]{Petrov2021}
{Petrov} L.,  2021, \mn@doi [\aj] {10.3847/1538-3881/abc4e1}, \href
  {https://ui.adsabs.harvard.edu/abs/2021AJ....161...14P} {161, 14}

\bibitem[\protect\citeauthoryear{{Plavin}, {Kovalev}, {Pushkarev}  \&
  {Lobanov}}{{Plavin} et~al.}{2019}]{Plavin2019}
{Plavin} A.~V.,  {Kovalev} Y.~Y.,  {Pushkarev} A.~B.,   {Lobanov} A.~P.,  2019,
  \mn@doi [\mnras] {10.1093/mnras/stz504}, \href
  {https://ui.adsabs.harvard.edu/abs/2019MNRAS.485.1822P} {485, 1822}

\bibitem[\protect\citeauthoryear{{Reid} \& {Honma}}{{Reid} \&
  {Honma}}{2014}]{Reid2014}
{Reid} M.~J.,  {Honma} M.,  2014, \mn@doi [\araa]
  {10.1146/annurev-astro-081913-040006}, \href
  {https://ui.adsabs.harvard.edu/abs/2014ARA&A..52..339R} {52, 339}

\bibitem[\protect\citeauthoryear{{Ricci} et~al.,}{{Ricci}
  et~al.}{2020}]{Ricci2020}
{Ricci} C.,  et~al., 2020, \mn@doi [\apjl] {10.3847/2041-8213/ab91a1}, \href
  {https://ui.adsabs.harvard.edu/abs/2020ApJ...898L...1R} {898, L1}

\bibitem[\protect\citeauthoryear{{Rioja} \& {Dodson}}{{Rioja} \&
  {Dodson}}{2020}]{Rioja2020}
{Rioja} M.~J.,  {Dodson} R.,  2020, \mn@doi [\aapr]
  {10.1007/s00159-020-00126-z}, \href
  {https://ui.adsabs.harvard.edu/abs/2020A&ARv..28....6R} {28, 6}

\bibitem[\protect\citeauthoryear{{Ruan}, {Anderson}, {Eracleous}, {Green},
  {Haggard}, {MacLeod}, {Runnoe}  \& {Sobolewska}}{{Ruan}
  et~al.}{2019a}]{Ruan2019B}
{Ruan} J.~J.,  {Anderson} S.~F.,  {Eracleous} M.,  {Green} P.~J.,  {Haggard}
  D.,  {MacLeod} C.~L.,  {Runnoe} J.~C.,   {Sobolewska} M.~A.,  2019a, arXiv
  e-prints, \href {https://ui.adsabs.harvard.edu/abs/2019arXiv190904676R} {p.
  arXiv:1909.04676}

\bibitem[\protect\citeauthoryear{{Ruan}, {Anderson}, {Eracleous}, {Green},
  {Haggard}, {MacLeod}, {Runnoe}  \& {Sobolewska}}{{Ruan}
  et~al.}{2019b}]{Ruan2019A}
{Ruan} J.~J.,  {Anderson} S.~F.,  {Eracleous} M.,  {Green} P.~J.,  {Haggard}
  D.,  {MacLeod} C.~L.,  {Runnoe} J.~C.,   {Sobolewska} M.~A.,  2019b, \mn@doi
  [\apj] {10.3847/1538-4357/ab3c1a}, \href
  {https://ui.adsabs.harvard.edu/abs/2019ApJ...883...76R} {883, 76}

\bibitem[\protect\citeauthoryear{{Runco} et~al.,}{{Runco}
  et~al.}{2016}]{Runco2016}
{Runco} J.~N.,  et~al., 2016, \mn@doi [\apj] {10.3847/0004-637X/821/1/33},
  \href {https://ui.adsabs.harvard.edu/abs/2016ApJ...821...33R} {821, 33}

\bibitem[\protect\citeauthoryear{{Shappee} et~al.,}{{Shappee}
  et~al.}{2014}]{Shappee2014}
{Shappee} B.~J.,  et~al., 2014, \mn@doi [\apj] {10.1088/0004-637X/788/1/48},
  \href {https://ui.adsabs.harvard.edu/abs/2014ApJ...788...48S} {788, 48}

\bibitem[\protect\citeauthoryear{{Sheng}, {Wang}, {Jiang}, {Yang}, {Yan}, {Dou}
   \& {Peng}}{{Sheng} et~al.}{2017}]{Sheng2017}
{Sheng} Z.,  {Wang} T.,  {Jiang} N.,  {Yang} C.,  {Yan} L.,  {Dou} L.,   {Peng}
  B.,  2017, \mn@doi [\apjl] {10.3847/2041-8213/aa85de}, \href
  {https://ui.adsabs.harvard.edu/abs/2017ApJ...846L...7S} {846, L7}

\bibitem[\protect\citeauthoryear{{Shepherd}, {Pearson}  \& {Taylor}}{{Shepherd}
  et~al.}{1994}]{Shepherd1994}
{Shepherd} M.~C.,  {Pearson} T.~J.,   {Taylor} G.~B.,  1994, \baas, \href
  {https://ui.adsabs.harvard.edu/abs/1994BAAS...26..987S} {26, 987}

\bibitem[\protect\citeauthoryear{{Szomoru}}{{Szomoru}}{2008}]{Szomoru2008}
{Szomoru} A.,  2008, in The role of VLBI in the Golden Age for Radio Astronomy.
  Sissa Medialab srl, p.~40

\bibitem[\protect\citeauthoryear{{Trakhtenbrot} et~al.,}{{Trakhtenbrot}
  et~al.}{2019}]{Trakhtenbrot2019}
{Trakhtenbrot} B.,  et~al., 2019, \mn@doi [\apj] {10.3847/1538-4357/ab39e4},
  \href {https://ui.adsabs.harvard.edu/abs/2019ApJ...883...94T} {883, 94}

\bibitem[\protect\citeauthoryear{{Ulvestad}}{{Ulvestad}}{2003}]{Ulvestad2003}
{Ulvestad} J.~S.,  2003, in {Zensus} J.~A.,  {Cohen} M.~H.,   {Ros} E.,  eds,
  Astronomical Society of the Pacific Conference Series Vol. 300, Radio
  Astronomy at the Fringe. Astron. Soc. Pac., p.~97 (\mn@eprint {arXiv}
  {astro-ph/0301057})

\bibitem[\protect\citeauthoryear{{Varenius} et~al.,}{{Varenius}
  et~al.}{2019}]{Varenius2019}
{Varenius} E.,  et~al., 2019, \mn@doi [\aap] {10.1051/0004-6361/201730631},
  \href {https://ui.adsabs.harvard.edu/abs/2019A&A...623A.173V} {623, A173}

\bibitem[\protect\citeauthoryear{{V{\'e}ron-Cetty} \&
  {V{\'e}ron}}{{V{\'e}ron-Cetty} \& {V{\'e}ron}}{2001}]{Veron-Cetty2001}
{V{\'e}ron-Cetty} M.~P.,  {V{\'e}ron} P.,  2001, \mn@doi [\aap]
  {10.1051/0004-6361:20010718}, \href
  {https://ui.adsabs.harvard.edu/abs/2001A&A...374...92V} {374, 92}

\bibitem[\protect\citeauthoryear{{Wadadekar}}{{Wadadekar}}{2004}]{Wadadekar2004}
{Wadadekar} Y.,  2004, \mn@doi [\aap] {10.1051/0004-6361:20034244}, \href
  {https://ui.adsabs.harvard.edu/abs/2004A&A...416...35W} {416, 35}

\bibitem[\protect\citeauthoryear{{Weiler}, {Panagia}, {Montes}  \&
  {Sramek}}{{Weiler} et~al.}{2002}]{Weiler2002}
{Weiler} K.~W.,  {Panagia} N.,  {Montes} M.~J.,   {Sramek} R.~A.,  2002,
  \mn@doi [\araa] {10.1146/annurev.astro.40.060401.093744}, \href
  {https://ui.adsabs.harvard.edu/abs/2002ARA&A..40..387W} {40, 387}

\bibitem[\protect\citeauthoryear{{Williams} et~al.,}{{Williams}
  et~al.}{2020}]{Williams2020}
{Williams} D.~R.~A.,  et~al., 2020, \mn@doi [\mnras] {10.1093/mnras/staa1152},
  \href {https://ui.adsabs.harvard.edu/abs/2020MNRAS.495.3079W} {495, 3079}

\bibitem[\protect\citeauthoryear{{Wo{\l}owska}, {Kunert-Bajraszewska}, {Mooley}
   \& {Hallinan}}{{Wo{\l}owska} et~al.}{2017}]{Woowska2017}
{Wo{\l}owska} A.,  {Kunert-Bajraszewska} M.,  {Mooley} K.,   {Hallinan} G.,
  2017, \mn@doi [Frontiers in Astronomy and Space Sciences]
  {10.3389/fspas.2017.00038}, \href
  {https://ui.adsabs.harvard.edu/abs/2017FrASS...4...38W} {4, 38}

\bibitem[\protect\citeauthoryear{{Yang}, {Wu}, {Paragi}  \& {An}}{{Yang}
  et~al.}{2012}]{Yang2012}
{Yang} J.,  {Wu} F.,  {Paragi} Z.,   {An} T.,  2012, \mn@doi [\mnras]
  {10.1111/j.1745-3933.2011.01182.x}, \href
  {https://ui.adsabs.harvard.edu/abs/2012MNRAS.419L..74Y} {419, L74}

\bibitem[\protect\citeauthoryear{{Yang}, {Paragi}, {Komossa}, {van Bemmel}  \&
  {Oonk}}{{Yang} et~al.}{2013}]{Yang2013}
{Yang} J.,  {Paragi} Z.,  {Komossa} S.,  {van Bemmel} I.,   {Oonk} R.,  2013,
  The Astronomer's Telegram, \href
  {https://ui.adsabs.harvard.edu/abs/2013ATel.5125....1Y} {5125, 1}

\bibitem[\protect\citeauthoryear{{Yang}, {Paragi}, {van der Horst}, {Gurvits},
  {Campbell}, {Giannios}, {An}  \& {Komossa}}{{Yang} et~al.}{2016}]{Yang2016}
{Yang} J.,  {Paragi} Z.,  {van der Horst} A.~J.,  {Gurvits} L.~I.,  {Campbell}
  R.~M.,  {Giannios} D.,  {An} T.,   {Komossa} S.,  2016, \mn@doi [\mnras]
  {10.1093/mnrasl/slw107}, \href
  {https://ui.adsabs.harvard.edu/abs/2016MNRAS.462L..66Y} {462, L66}

\bibitem[\protect\citeauthoryear{{Yang} et~al.,}{{Yang}
  et~al.}{2018}]{Yang2018}
{Yang} Q.,  et~al., 2018, \mn@doi [\apj] {10.3847/1538-4357/aaca3a}, \href
  {https://ui.adsabs.harvard.edu/abs/2018ApJ...862..109Y} {862, 109}

\bibitem[\protect\citeauthoryear{{Yang}, {Gurvits}, {Paragi}, {Frey}, {Conway},
  {Liu}  \& {Cui}}{{Yang} et~al.}{2020}]{Yang2020imbh}
{Yang} J.,  {Gurvits} L.~I.,  {Paragi} Z.,  {Frey} S.,  {Conway} J.~E.,  {Liu}
  X.,   {Cui} L.,  2020, \mn@doi [\mnras] {10.1093/mnrasl/slaa052}, \href
  {https://ui.adsabs.harvard.edu/abs/2020MNRAS.495L..71Y} {495, L71}

\bibitem[\protect\citeauthoryear{{Yang}, {Paragi}, {Nardini}, {Baan}, {Fan},
  {Mohan}, {Varenius}  \& {An}}{{Yang} et~al.}{2021a}]{Yang2021pds}
{Yang} J.,  {Paragi} Z.,  {Nardini} E.,  {Baan} W.~A.,  {Fan} L.,  {Mohan} P.,
  {Varenius} E.,   {An} T.,  2021a, \mn@doi [\mnras] {10.1093/mnras/staa2445},
  \href {https://ui.adsabs.harvard.edu/abs/2021MNRAS.500.2620Y} {500, 2620}

\bibitem[\protect\citeauthoryear{Yang et~al.,}{Yang et~al.}{2021b}]{Yang2021}
Yang J.,  et~al., 2021b, \mn@doi [\mnras] {10.1093/mnrasl/slab005}, 502, L61

\bibitem[\protect\citeauthoryear{{Yuan} \& {Narayan}}{{Yuan} \&
  {Narayan}}{2014}]{Yuan2014}
{Yuan} F.,  {Narayan} R.,  2014, \mn@doi [\araa]
  {10.1146/annurev-astro-082812-141003}, \href
  {https://ui.adsabs.harvard.edu/abs/2014ARA&A..52..529Y} {52, 529}

\makeatother
\end{thebibliography}




\appendix

\section{Additional images} 

\begin{figure}
\centering
\includegraphics[width=0.48\textwidth]{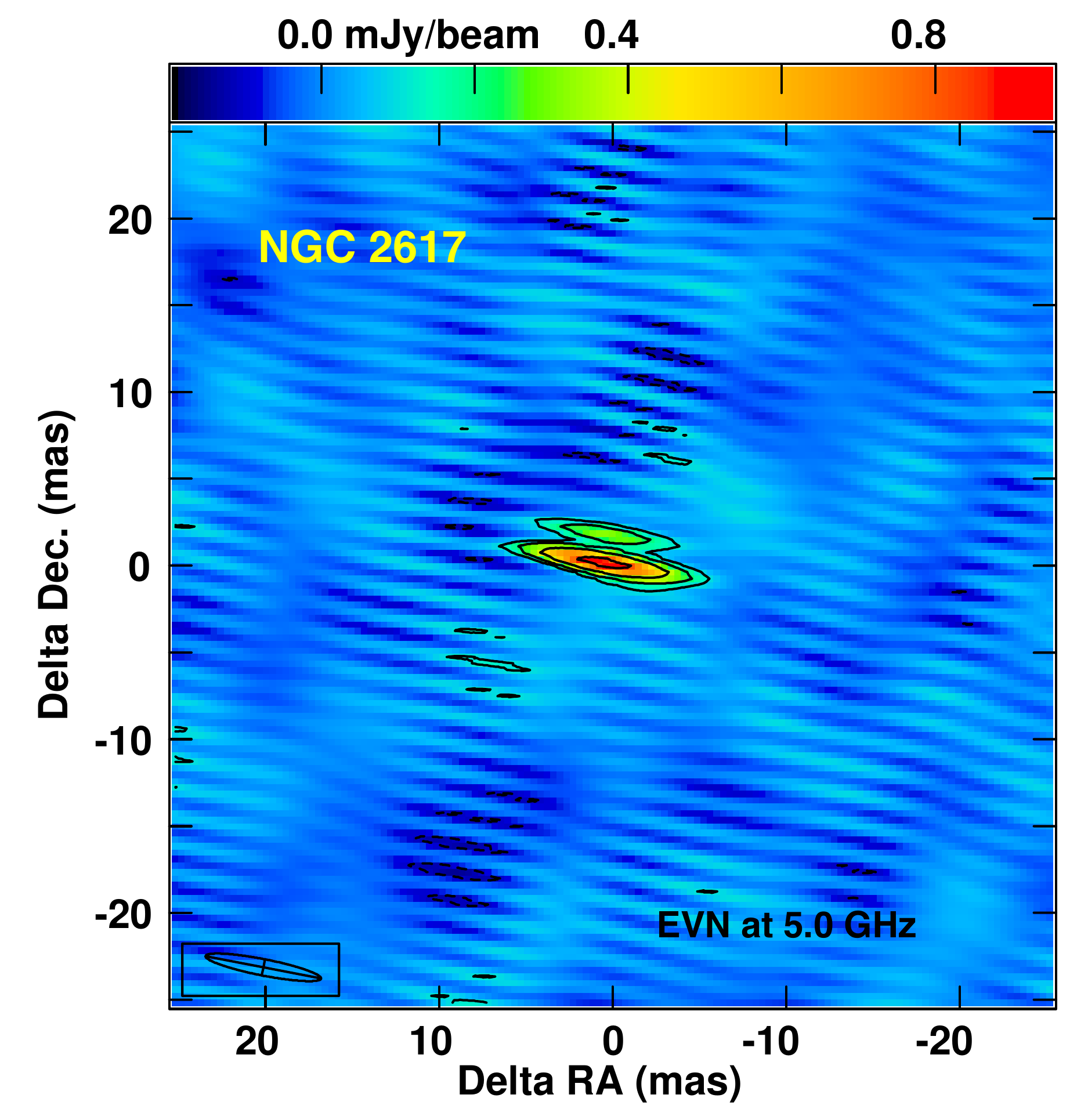}  \\
\caption{
Hint for an inner jet component in the nearby changing-look Seyfert galaxy NGC~2617 observed by the EVN at 5~GHz with the uniform grid weighting on 2013 September 18. The contours are at the levels 2.5$\sigma$~$\times$~($-1$, 1, 2, 4, 8) and $\sigma$~= 0.041~mJy\,beam$^{-1}$. The map peak bright is 0.95~mJy\,beam$^{-1}$. The beam FWHM is 6.76~$\times$~0.95~mas$^{2}$ at $78\fdg9$.}
\label{fig:evn_c_uw}
\end{figure}

\begin{figure*}
\centering
\includegraphics[width=0.48\textwidth]{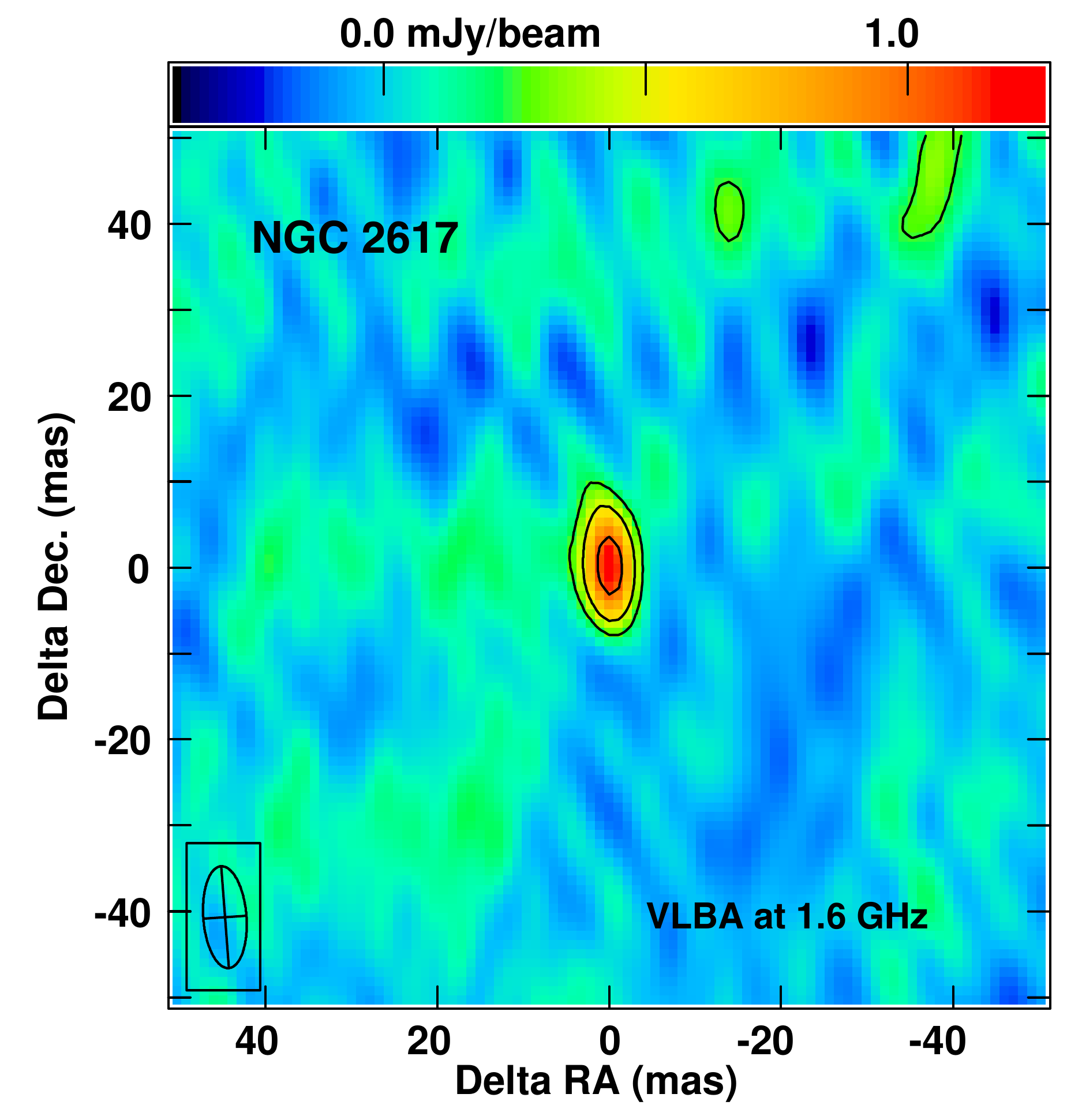}  
\includegraphics[width=0.48\textwidth]{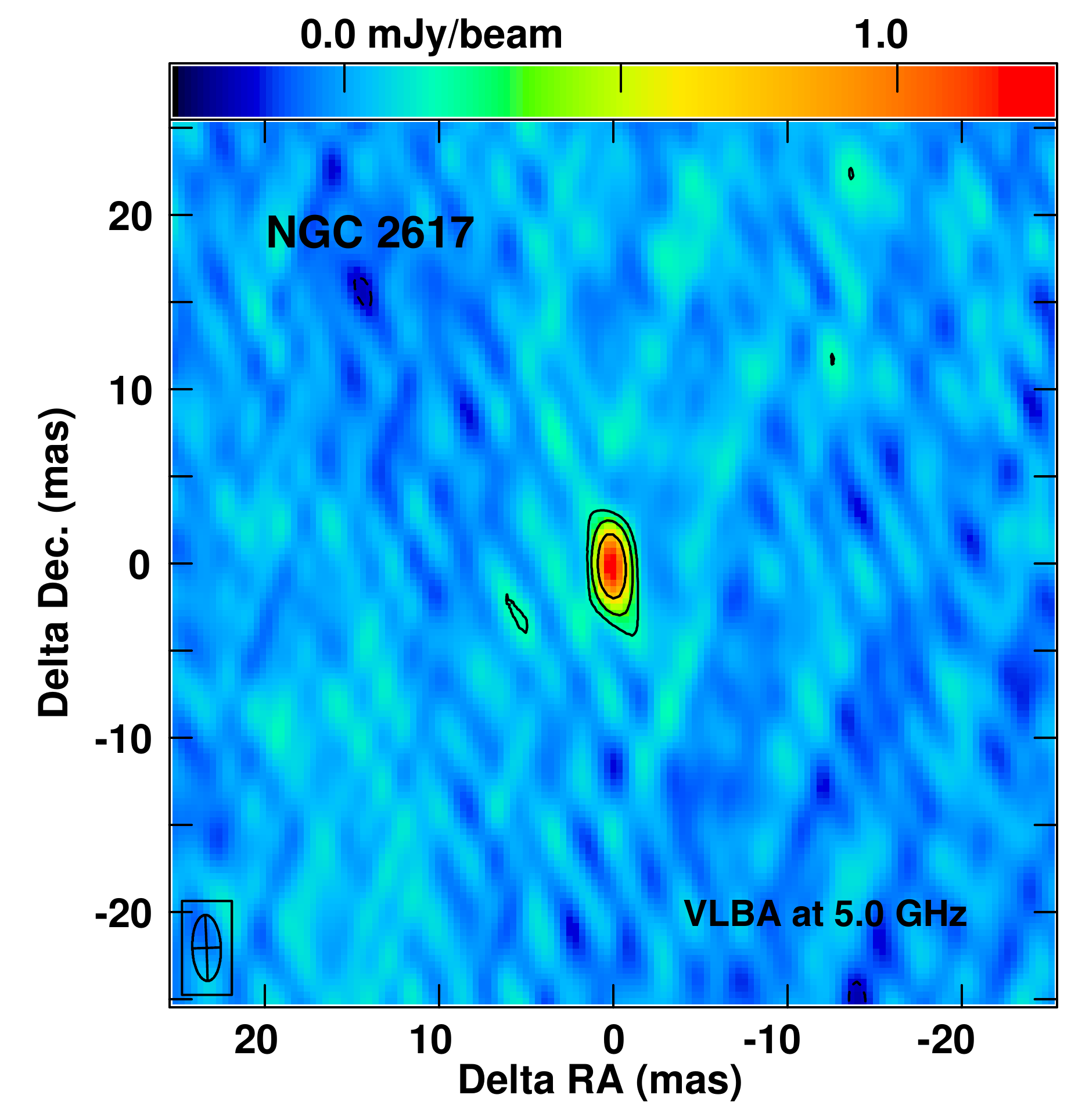}  \\
\caption{
The compact radio morphology of the nearby changing-look Seyfert galaxy NGC~2617 observed by the VLBA at 1.6 and 5~GHz. The first contours are at the level of 2.5$\sigma$. The images were made with natural grid weighting. The map parameters are reported in Table~\ref{tab:flux}. }
\label{fig:vlba}
\end{figure*}

\begin{figure*}
\centering
\includegraphics[width=0.48\textwidth]{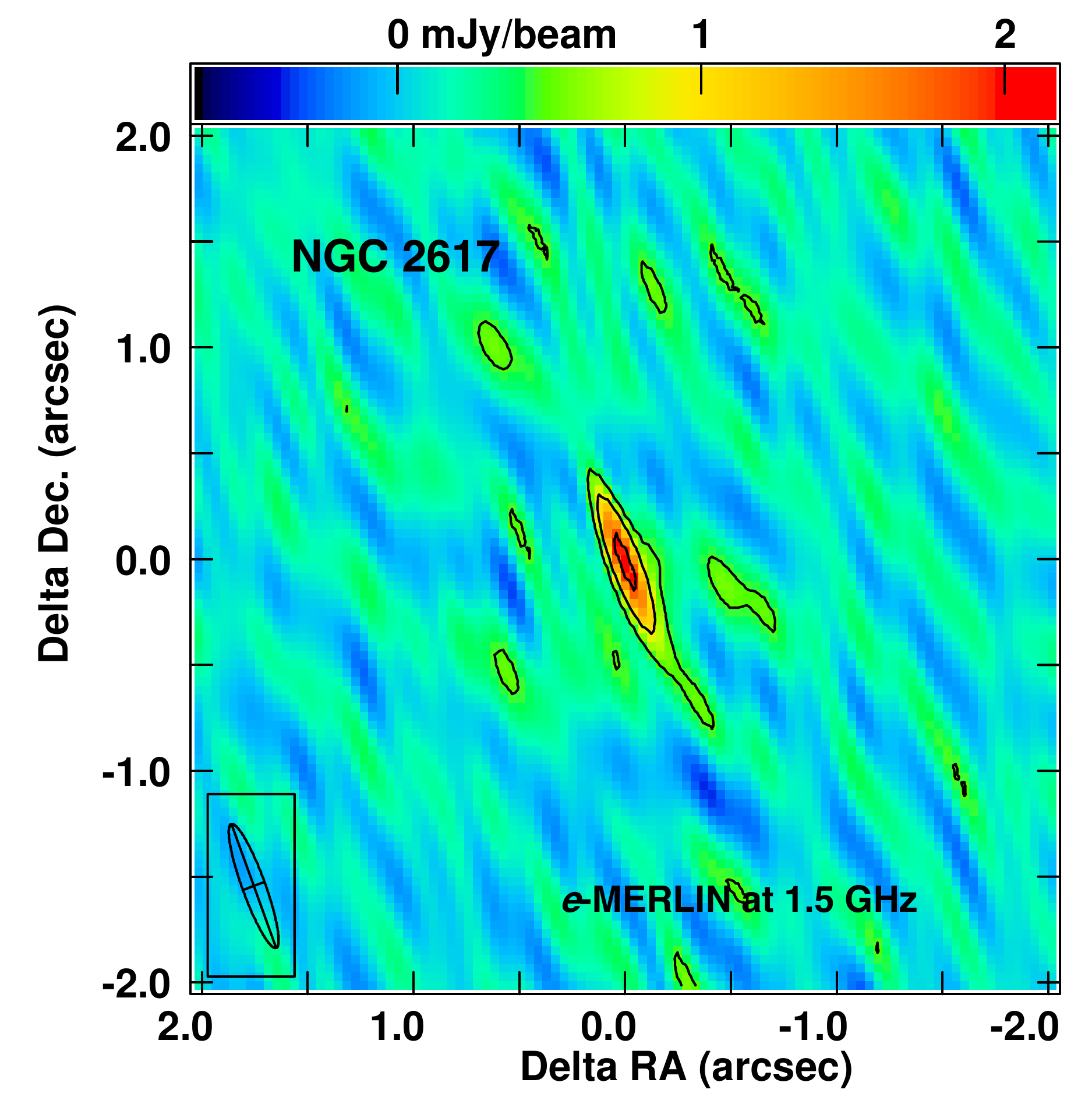}  
\includegraphics[width=0.48\textwidth]{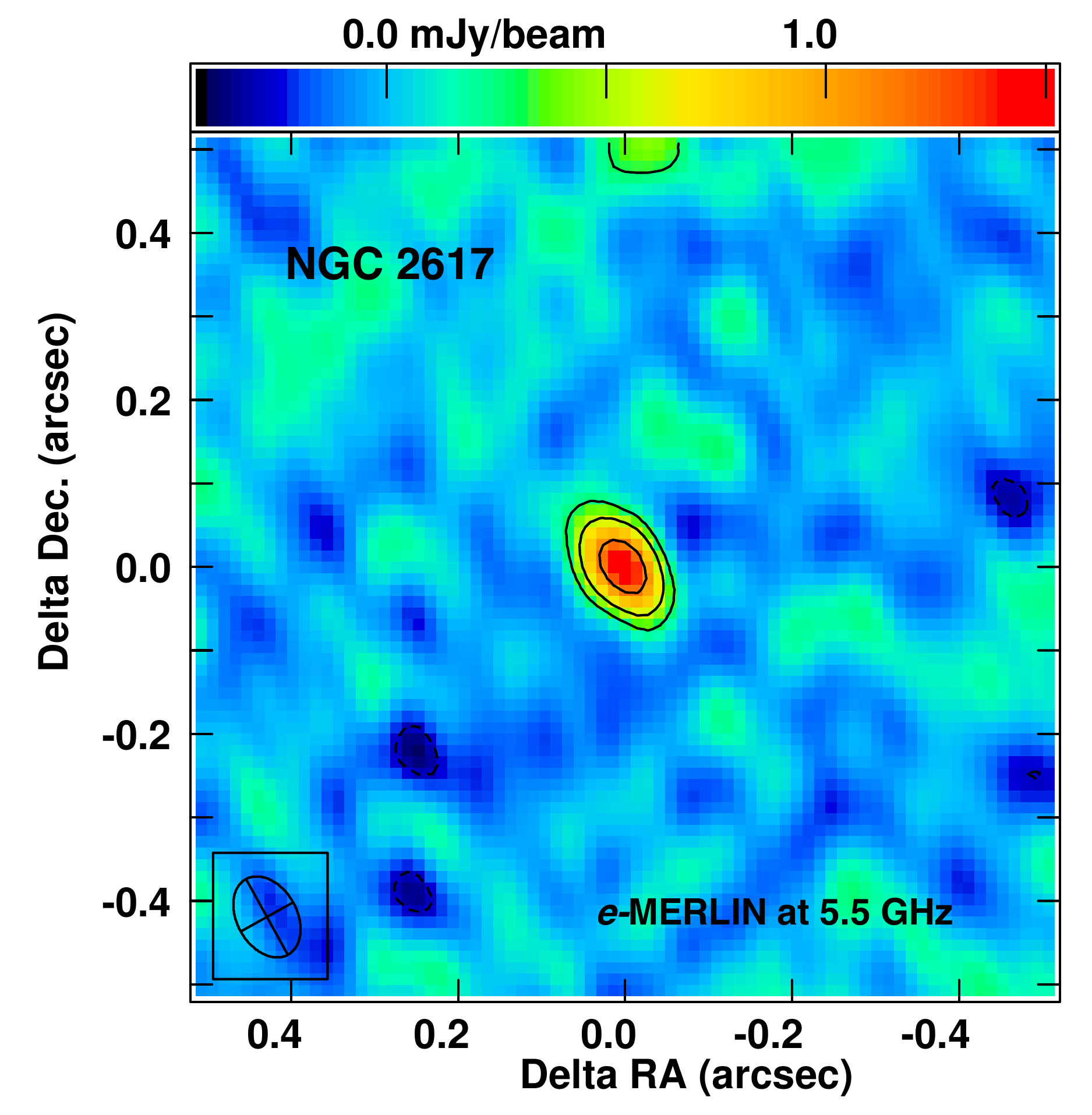}  \\
\caption{
The compact radio structure of NGC~2617 observed by the \textit{e}-MERLIN at 1.5 and 5.5~GHz. The first contours are at the level of 2.5$\sigma$. The images were made with uniform grid weighting at 1.5~GHz and natural grid weighting at 5.5~GHz. The map parameters are reported in Table~\ref{tab:flux}. }
\label{fig:emerlin}
\end{figure*}



\bsp	
\label{lastpage}
\end{document}